\documentclass[12pt]{article}

\usepackage{scicite}
\usepackage{graphicx}

\usepackage{times}

\usepackage{longtable}
\usepackage{calrsfs}
\DeclareMathAlphabet{\pazocal}{OMS}{zplm}{m}{n}
\usepackage{amsmath}
\usepackage{amssymb}
\usepackage{hyperref}

\newcommand{\au}{\, {\rm au}}
\newcommand{\msunyr}{\, {\rm M}_{\odot}\,{\rm yr^{-1}}}

\newenvironment{sciabstract}{%
\begin{quote} \bf}
{\end{quote}}

\title{\vspace{-5cm}\vspace*{\fill}Radial-velocity discovery of a second planet in the TOI-1338/BEBOP-1 circumbinary system \\(Old title: The First Circumbinary Planet Discovered with Radial Velocities)\vspace{-2cm}\footnote{The Version of Record of this article is published in Nature Astronomy, and is available online at \url{https://doi.org/10.1038/s41550-023-01948-4}}}

\date{}

\begin{document} 

% Double-space the manuscript.

% Make the title.

\maketitle 

\begin{center}

\author
{\textbf{Matthew R. Standing$^{1,2\ast}$,
%mxs1263@bham.ac.uk
Lalitha Sairam$^{1}$,
%L.Sairam@bham.ac.uk
David V. Martin$^{3,4}$,\\ 
%martin.4096@osu.edu
Amaury H. M. J. Triaud$^{1}$, 
%a.triaud@bham.ac.uk
Alexandre C.~M. Correia$^{5,6}$, 
%acor@uc.pt
Gavin A. L. Coleman$^{7}$,\\
%gavin.coleman@qmul.ac.uk
Thomas A. Baycroft$^{1}$,
%txb187@student.bham.ac.uk
Vedad Kunovac$^{8,9}$, 
%vedad.kunovac@lowell.edu
Isabelle Boisse$^{10}$,
%isabelle.BOISSE@univ-amu.fr
Andrew Collier Cameron$^{11}$, \\
%acc4@st-andrews.ac.uk
Georgina Dransfield$^{1}$,
Jo\~ao P. Faria$^{12,13}$, 
%joao.faria@astro.up.pt
Micha\"el Gillon$^{14}$, 
%michael.gillon@uliege.be
Nathan C. Hara$^{15}$, \\
%Nathan.Hara@unige.ch
Coel Hellier$^{16}$,
%c.hellier@keele.ac.uk
Jonathan Howard$^{1}$,
%JXH1104@student.bham.ac.uk
Ellie Lane$^{1}$,
%eml873@student.bham.ac.uk
Rosemary Mardling$^{17}$,\\
%rosemary.mardling@monash.edu
Pierre F. L. Maxted$^{16}$, 
%p.maxted@keele.ac.uk
Nicola J. Miller$^{16}$,
%n.j.miller1@keele.ac.uk
Richard P. Nelson$^{7}$,\\ 
%r.p.nelson@qmul.ac.uk
Jerome A. Orosz$^{18}$, 
%jorosz@sdsu.edu
Franscesco Pepe$^{15}$, 
%francesco.pepe@unige.ch
Alexandre Santerne$^{10}$,\\
%alexandre.santerne@lam.fr
Daniel Sebastian$^{1}$,
%D.Sebastian.1@bham.ac.uk
St\'ephane Udry$^{15}$,
%stephane.udry@unige.ch
William F. Welsh$^{18}$}\\
%wwelsh@sdsu.edu

\normalsize{$^{1}$School of Physics and Astronomy, University of Birmingham, Edgbaston,}\\
\normalsize{Birmingham B15 2TT, UK}\\
\normalsize{$^{2}$School of Physical Sciences, The Open University, Milton Keynes, MK7 6AA, UK}\\
\normalsize{$^{3}$Department of Astronomy, The Ohio State University, 4055 McPherson Laboratory,}\\
\normalsize{Columbus, OH 43210, USA}, \normalsize{$^{4}$NASA Sagan Fellow}\\
\normalsize{$^{5}$CFisUC, Departamento de F\'isica, Universidade de Coimbra, 3004-516 Coimbra, Portugal}\\
\normalsize{$^{6}$IMCCE, UMR8028 CNRS, Observatoire de Paris, PSL Universit\'e,}\\
\normalsize{77 av. Denfert-Rochereau,75014 Paris, France}\\
\normalsize{$^{7}$Astronomy Unit, Queen Mary University of London, Mile End Road, London E1 4NS, UK}\\
\normalsize{$^{8}$Lowell Observatory, 1400 W. Mars Hill Rd., Flagstaff, AZ 86001, USA}\\
\normalsize{$^{9}$Department of Astronomy and Planetary Science, Northern Arizona University, }\\
\normalsize{PO Box 6010, Flagstaff, AZ 86011}\\
\normalsize{$^{10}$Aix Marseille Univ, CNRS, CNES, Institut Origines, LAM, Marseille, France}\\
\normalsize{$^{11}$ Centre for Exoplanet Science / SUPA, School of Physics and Astronomy, }\\ \normalsize{University of St Andrews, North Haugh, St Andrews, Fife, KY16 9SS, UK}\\
\normalsize{$^{12}$ Instituto de Astrof\'iısica e Ci\^encias do Espa\c co, Universidade do Porto, }\\\normalsize{CAUP, Rua das Estrelas, 4150-762 Porto, Portugal}\\
\normalsize{$^{13}$ Departamento de F\'isica e Astronomia, Faculdade de Ci\^encias, Universidade do Porto, }\\\normalsize{Rua do Campo Alegre, 4169-007 Porto, Portugal}\\
\normalsize{$^{14}$ Astrobiology Research Unit, University of Li\`ege, All\'ee du 6 ao\^ut 19 (B5C), }\\\normalsize{4000 Li\`ege (Sart-Timan), Belgium}\\
\normalsize{$^{15}$ Observatoire Astronomique de l’Universit\'e de Gen\`eve, }\\\normalsize{Chemin de Pegasi 51, 1290 Versoix, Switzerland}\\
\normalsize{$^{16}$ Astrophysics Group, Keele University, ST5 5BG, UK}\\
\normalsize{$^{17}$ School of Physics and Astronomy, Monash University, Victoria, 3800, Australia}\\
\normalsize{$^{18}$ Department of Astronomy, San Diego State University, 5500 Campanile Drive, }\\\normalsize{San Diego, CA 92182, USA}\\

\normalsize{$^\ast$E-mail: matthew.standing@open.ac.uk}
}
    
\end{center}
% Place your abstract within the special {sciabstract} environment.
\baselineskip24pt

\begin{sciabstract}
    We report the detection of a gas-giant planet in orbit around both stars of an eclipsing binary star system that also contains the smaller, inner transiting planet TOI-1338b.
    The new planet, called TOI-1338/BEBOP-1c, was discovered using radial-velocity data collected with the HARPS and ESPRESSO spectrographs. Our analysis reveals it is a $65.2~\rm{M_{\oplus}}$  circumbinary planet with a period of $215.5$~days.
    This is the first detection of a circumbinary planet using radial-velocity observations alone, and makes TOI-1338/BEBOP-1 only the second confirmed multiplanet circumbinary system to date.
    We do not detect the smaller inner transiting planet with radial-velocity data, and can place an upper limit on the inner planet's mass at $21.8~\mathrm{M}_\oplus$ with $99\%$ confidence. The inner planet is the first circumbinary planet amenable for atmospheric characterisation, using the James Webb Space Telescope.
\end{sciabstract}

% In setting up this template for *Science* papers, we've used both
% the \section* command and the \paragraph* command for topical
% divisions.  Which you use will of course depend on the type of paper
% you're writing.  Review Articles tend to have displayed headings, for
% which \section* is more appropriate; Research Articles, when they have
% formal topical divisions at all, tend to signal them with bold text
% that runs into the paragraph, for which \paragraph* is the right
% choice.  Either way, use the asterisk (*) modifier, as shown, to
% suppress numbering.

%\section*{Introduction}

Circumbinary planets are planets that orbit both stars of a central binary. They were once confined to science fiction, but the discovery of Kepler-16b \cite{Doyle2011} paved the way for the discovery of 14 transiting planets in 12 binary systems, by the \textit{Kepler} \cite{Borucki2010a} and \textit{TESS} \cite{Ricker2015} missions. 
Of the 12 transiting circumbinary planet systems discovered to date only one hosts multiple circumbinary planets, Kepler-47. Kepler-47 b, d and c have orbital periods of $49.5$, $187.4$, and $303.2~\rm{days}$ respectively \cite{Orosz2019}, with c's orbit placing it within the system's habitable zone.
%Only two of the three planets in the Kepler-47 system posses a mass determined to be different from zero by $>2\sigma$, both obtained from planet transit timing variations: Kepler-47~d with a mass of $19.02^{+23.84}_{-11.67}~\rm M_\oplus$, and the smallest outer planet Kepler-47~c, with a mass of $3.17^{+2.18}_{-1.25}~\rm M_\oplus$.
Planet forming disks around binaries are harsh environments for planet formation to take place, and circumbinary planet discoveries provide insights into the formation and migration mechanisms at play in these unique environments \cite{Meschiari2012, Lines2014, Pierens2020, Martin2022,Fitzmaurice2022}.

Of the 14 confirmed transiting circumbinary planets discovered to date, there are only significant mass detections (different from zero at $>2\sigma$) for six of them. These are: Kepler-34~b and 35~b \cite{Welsh2012}, TIC 17290098~b \cite{Kostov2021}, Kepler-16~b \cite{Doyle2011, Triaud2022}, Kepler-47~c and d \cite{Orosz2019}. Where the masses were determined from binary eclipse timing variations (ETV's) alone.
For the remaining eight circumbinary planets only upper limits can be placed on their masses, because the ETV's are on the order of only seconds or minutes, and hence difficult to measure. Their masses could therefore be much lower than expected and reveal several inflated objects, ideal targets for atmospheric transmission follow-up observations \cite{Kempton2018}. To correctly characterise these planets, accurate masses are required.

To increase the number of known circumbinary planets, and to provide accurate masses for systems discovered with the transit method, we initiated a radial-velocity observing survey dedicated to circumbinary planet detection called Binaries Escorted By Orbiting Planets (BEBOP) \cite{Martin2019}. Systems followed by BEBOP are on average 4 visual magnitudes brighter than circumbinary systems identified with {\it Kepler} \cite{Martin2019}. 
Radial velocities are less restricted to the edge-on and shorter orbital periods found by the transit method \cite{Martin&Triaud2014}. Despite initial challenges in bringing radial-velocity precision for binaries down to values where planets can be detected \cite{Konacki2009b}, recent results have produced an independent detection of Kepler-16b with a precision of $1.5~\rm m.s^{-1}$, resulting in a planetary mass in agreement with ETV measurements \cite{Triaud2022}.

The first circumbinary planet discovered by NASA's Transiting Exoplanet Survey Satellite (TESS) \cite{Ricker2015} was TOI-1338~b \cite{Kostov2020}.
TOI-1338 is a low mass eclipsing binary system, which consists of a $1.13~\rm M_\odot$ F8-type primary star with a visual magnitude of $V = 11.7$ \cite{Kostov2020, Triaud2017}, and a $0.31~\rm M_\odot$ M-dwarf companion.
This system was already being monitored with radial velocities with HARPS \cite{Pepe2002b} as part of the BEBOP project, where the target was known as EBLM~J0608-59 \cite{Triaud2017,Martin2019}. 
Radial-velocity data available at the time were unable to detect any trace of TOI-1338b, %the $33 \pm 20 ~\rm M_{\oplus}$, 
the Saturn-sized planet announced by Kostov et al. (2020) \cite{Kostov2020}.
To constrain the mass of this planet, we used the ESPRESSO spectrograph at the Very Large Telescope \cite{Pepe2021}. Meanwhile we continued to observe with HARPS intermittently in order to combine both datasets more easily, and in order to mitigate observatory closures caused by the COVID pandemic. In total, we collected $123$ ESPRESSO, and $61$ HARPS spectra covering a timespan of $1472$~days. 

We analyse our radial-velocity data with the diffusive nested sampler implemented in \texttt{kima} \cite{Faria2018}, slightly adapted for the purpose of a circumbinary system. We include non Keplerian effects such as gravitational redshift, light time travel \cite{Zucker2007, Konacki2010, Sybilski2013} and tidal distortion \cite{Arras2012}. In addition, we directly fit for the most important of the Newtonian perturbations, the apsidal precession, which we parameterise by a free parameter, $\dot{\omega}_{\rm bin}$. Further details can be found in the supplementary material, but we note here that a particularly useful feature of \texttt{kima} is that the number of planetary signals in the data can be included as a free parameter in the analysis. The number of planets detected is determined from the ratio of evidence between models with different number of planets. The radial-velocity data favours a one-planet model over a zero-planet model with a ratio of probability (Bayes factor) $>29,000$ (where $>150$ is the typical threshold for detection).
The same analysis provides a Bayes factor of only $1.4$ in favour of a two-planet model over a one-planet model, far below the detection threshold. Therefore, only one planetary signal is formally detected in our data.

Figure~1 depicts a histogram of the posterior samples obtained from our analysis. There are two peaks, one at the binary orbital period of $14.6$~days, and a single additional excess at $215.5$~days. Since this periodicity does not correspond to the $95$~day period of the transiting planet \cite{Kostov2020}, and an outer orbit significantly improves the fit, we conclude the detected planetary signal is likely that of an additional, outer circumbinary planet within the system. 
We therefore give the system the name BEBOP-1, becoming the first entry in the BEBOP catalogue for circumbinary planets detected with radial velocities, with the new signal being TOI-1338/BEBOP-1c\footnote{The name BEBOP is recognised as an official designation by the IAU, and indicates the availability of high-precision radial velocity measurements on the target and their ability to detect a planet.}.
Figure~2 shows a phase-folded plot of the radial-velocity variation caused by TOI-1338/BEBOP-1c, with a semi-amplitude $K_{\rm{c}}=5.6\pm 1.0~\rm{m.s^{-1}}$, corresponding to a minimum mass $m_{\rm c} \sin i_{\rm c}=0.217\pm 0.035~\rm M_{Jup}$. The binary's apsidal precession is found marginally different from zero ($\dot{\omega}_{\rm bin} = 66.0^{+65.0}_{-54.7}~\rm{arcsec\,year^{-1}}$). Parameters of the binary and planetary orbits, obtained from our model fit to the radial velocities are available in Table~1. 
We calculate the True Inclusion Probability (TIP) \cite{Hara2021, Hara2022} for our posterior samples which also provides a clear $>99.9\%$ probability of the presence of a planetary signal at a period of $215.5$~days. Further details can be found in the supplementary material.

Stellar magnetic activity can produce periodic modulations in radial velocity over a wide range of timescales sometimes mimicking a planetary companion \cite{Queloz2001}. To verify if this is the case for the TOI-1338/BEBOP-1c signal, we search for periodic signals in five spectroscopic activity indices (see supplementary material for details). 

Stellar activity has been shown to affect most, if not all of these five indices at once \cite{Gomes2011}. In our case, only two show any signal. Furthermore, at shorter periods neither Ca II H+K nor H$\alpha$ indices produce any significant peak. Combining the stellar radius and projected rotational velocity of TOI-1338/BEBOP-1, a rotation period of $19\pm3$~days has been reported \cite{Kostov2020}. We do not find any statistically significant signal at the estimated rotation period. 

Our analysis of activity indices rules out any association between the 215~day radial-velocity variation and stellar activity, supporting the hypothesis for TOI-1338/BEBOP-1c's planetary nature. Further details and figures are provided in the supplementary material. 

%Stellar magnetic activity can produce periodic modulations in radial velocity over a wide range of timescales sometimes mimicking a planetary companion \cite{Queloz2001}. 
%To verify whether this is the case for the signal we attach to BEBOP-1c, we search for periodic signals in five spectroscopic activity indices (see supplementary material for details). Stellar activity has been shown to affect if not all, most of these five indices at once \cite{Gomes2011}, however, in our case only two show any signal. 

%At shorter periods, the  Ca II H + K index does not produce any significant peak with periodicity whereas, the H$\alpha$ index shows a significant peak around $16$ days. 

%Combining the stellar radius and projected rotational velocity of  BEBOP-1's rotation periods of $19\pm3$~days have been reported \cite{Kostov2020}. We interpret the $16$ days periodicity as tracing the rotation period of BEBOP-1A. The binary orbital period is slightly shorter at 14.6 days. Given the small mass ratio ($M_2/M_1=0.28$) it is not surprising that the primary star does not rotate synchronously with the binary star orbit \cite{Lurie2017}. 

%Additionally, the H$\alpha$ index suggests the existence of a long term stellar signal ($\sim 280~\rm days$), possibly associated to magnetic cycles like observed in Kepler-16 \cite{Triaud2022}. Our analysis of activity indices rules out any association between the 215~day radial velocity variation and stellar activity near, confirming BEBOP-1c's planetary nature. Further details and figures are provided in the supplementary material. 

A visual inspection of \textit{TESS} lightcurves shows no transit of TOI-1338/BEBOP-1c, however, thanks to orbital circulation, transits are expected to occur in due time. Circumbinary orbits exhibit nodal precession. This changes the orientation of a circumbinary planet's orbital plane with respect to both the binary and the observer. This makes a planet change from a transiting to a non-transiting configuration \cite{Schneider1994,Martin&Triaud2014} as has been seen in a few systems \cite{Kostov2014,Welsh2015}. 
Using an analytic criterion  \cite{Martin&Triaud2015,Martin2017b},
we find that TOI-1338/BEBOP-1c is {\it guaranteed to eventually transit} mainly because the binary is so well-aligned with our line of sight ($I_{\rm bin}=89.658^{\circ}$) combined with the rather large size of the primary star ($R_{\rm A}=1.299R_{\odot}$). Whilst TOI-1338/BEBOP-1c will {\it eventually} transit, we are unable to predict when and how frequently. Its precession period is of order 119 years, during which time there will be two periods of transitability of a duration depending on TOI-1338/BEBOP-1c's orbital inclination.% The fact that BEBOP-1c has not transited places a constraint on its inclination (David checking) 

% While we are confident the radial-velocity signal we detect is that of a circumbinary planet, it is also worth verifying whether the system as we propose it is plausible. 
To verify the orbital parameters of our fit we carry out a global stability analysis of the system. This stability analysis shows the system is stable when the two circumbinary planets are in nearly circular and coplanar orbits (see supplementary material). More precisely, the eccentricities of both planets cannot exceed 0.1, in agreement with the best fit solution (Table~\ref{tab:results}). The inclination of planet c, $i_{\rm c}$, is unconstrained by the radial velocities, however, thanks to orbital stability arguments, it cannot be higher than $40^\circ$ with respect to the orbital plane of the binary, corresponding to a maximum mass of $m_{\rm max, c} = 0.28~\rm M_{Jup}$.

\begin{figure}
    \centering
	\includegraphics{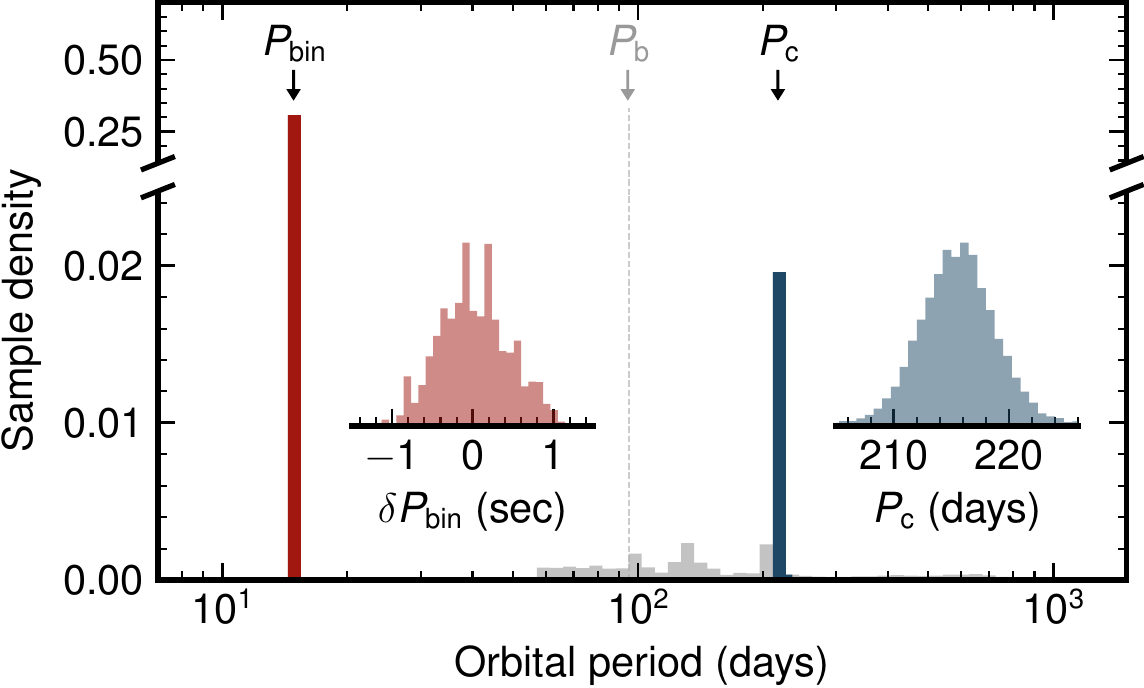}
    \caption{Histogram periodogram of posterior samples obtained from a \texttt{kima} run on the TOI-1338/BEBOP-1 system. The $14.6$~day binary period along with the $215.5$~day period of TOI-1338/BEBOP-1c can be clearly seen, and are highlighted in red and blue respectively. The inlaid plots show a zoom on these two peaks. No significant peak from the transiting planet TOI-1338/BEBOP-1b can be seen in these posterior samples.}
    \label{fig:BEBOP-1_periodogram}
\end{figure}

\begin{figure}
    \centering
	\includegraphics{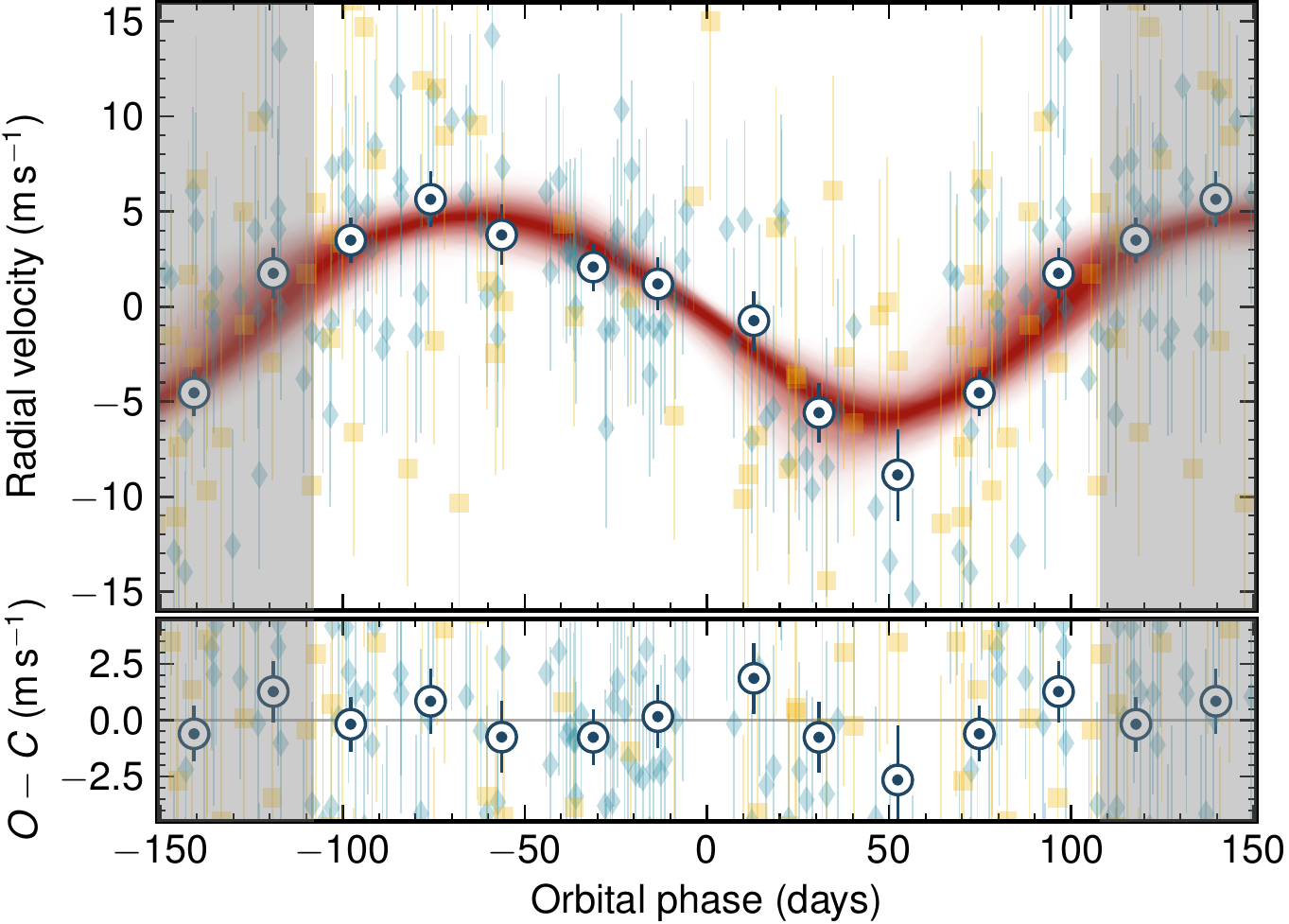}
    \caption{Phased Keplerian Radial-Velocity (RV) models of TOI-1338/BEBOP-1c with ESPRESSO (blue diamonds) and HARPS (orange squares) data along with associated residuals after removing the binary signal. RV data is binned by $0.1$ phase units (${\sim}21.6$~days) and illustrated by the circular points.
    Red Keplerian models are based on $500$ randomly drawn posterior samples from a \texttt{kima} run, shaded from the $50^{\rm{th}}$ to $99^{\rm{th}}$ percentiles. The shaded regions display the repeating signal.}
    \label{fig:BEBOP-1_phased}
\end{figure}

\begin{figure}
    \centering
	\includegraphics[width=\columnwidth]{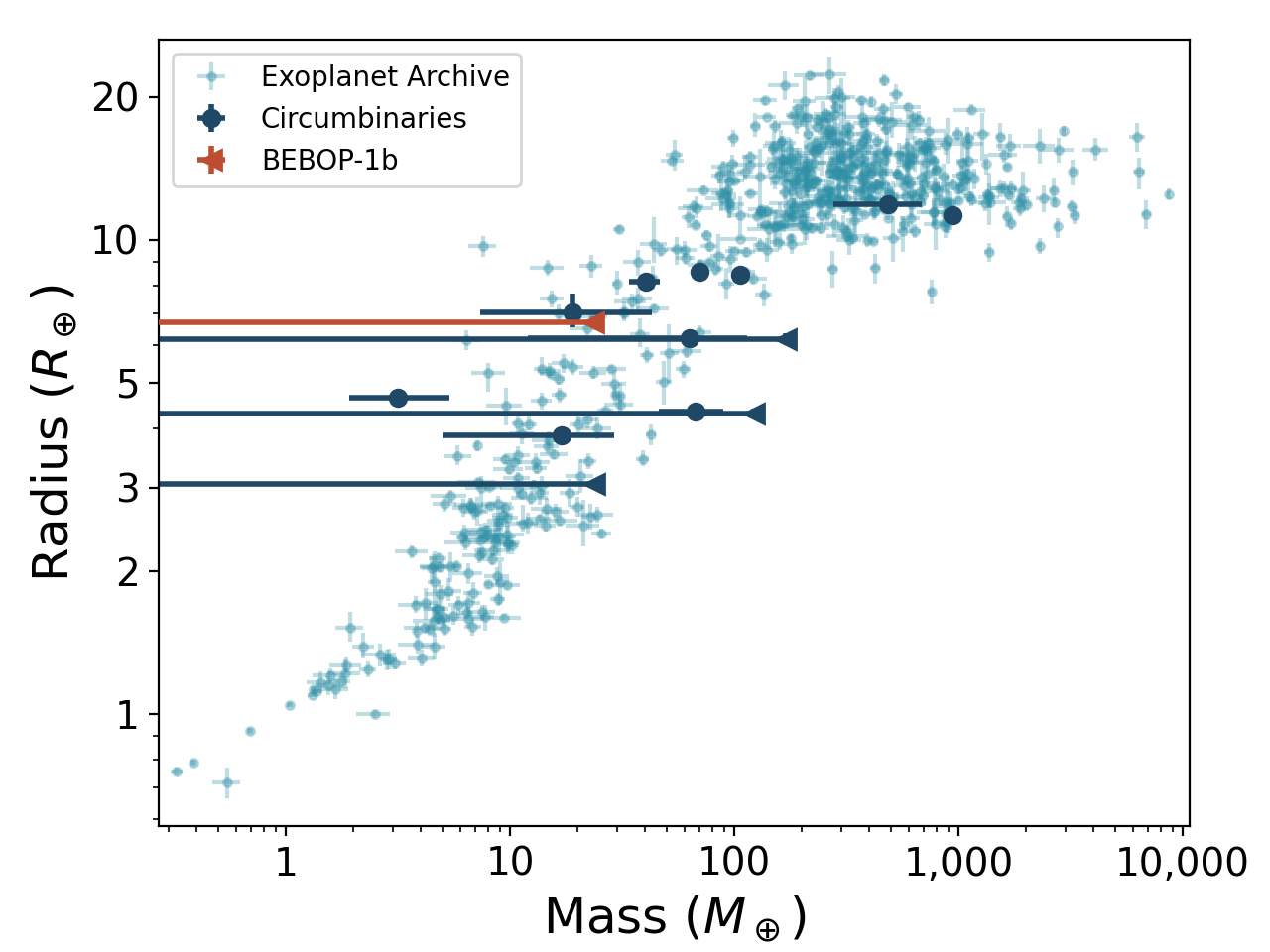}
    \caption{Radius vs Mass plot of all transiting circumbinary planets (dark blue) and planets orbiting single stars (light blue). TOI-1338/BEBOP-1b is highlighted in orange, with one of the lowest densities known. The newly-discovered TOI-1338/BEBOP-1c is not on the graph since it does not transit. For the single star planets we restrict the sample to planets with mass and radius percentage errors less than $20\%$.}
    \label{fig:mass_radius}
\end{figure}

The influence of the central binary means that circumbinary planets have unique formation pathways, and it is noteworthy that the system discussed in this paper is only the second system known to host multiple planets. Hydrodynamical simulations have shown that the preference for observed circumbinary planets to be located close to the stability boundary can be explained by formation at large distances in circumbinary discs, followed by inwards migration and stalling at the edge of the inner cavities formed by the central binaries \cite{Pierens07,Pierens08,Penzlin22}. 
We have utilised a purpose-built simulation code to study the formation of circumbinary planets as a means of understanding plausible formation scenarios for the TOI-1338/BEBOP-1 system. Details of this code are available in the supplementary material. The model includes the N-body integrator \textsc{mercury6}, adapted to include a  central binary system \cite{Chambers99,Chambers02}, and it incorporates prescriptions for a viscous circumbinary disc that includes the effects of an eccentric, precessing central cavity and photoevaporative winds \cite{Coleman22}, pebble accretion onto planetary seeds \cite{Lambrechts14}, planet migration \cite{Paardekooper11}, and gas accretion onto growing planets \cite{Coleman17, Poon21}. 
Our suite of simulations produced numerous systems that were qualitatively similar to TOI-1338/BEBOP-1 (see additional material), with TOI-1338/BEBOP-1b and c analogues landing on stable orbits at their observed locations.

With no detection of the $95$~day planet in our data the best we can do is to calculate a detection limit for this period region. We find that TOI-1338/BEBOP-1b has a mass  $<~21.8 \rm~M_\oplus$ with $99\%$ confidence. This is compatible with the $33\pm20~\rm M_\oplus$ from ETV's \cite{Kostov2020}. Combined with a planetary radius of $\approx6.9~\rm R_\oplus$ \cite{Kostov2020}, TOI-1338/BEBOP-1b has a maximum mean planetary density $<0.36~\rm{g.cm^{-3}}$. This allows us to calculate the Transmission Spectroscopy Metric (TSM) \cite{Kempton2018} for planet b using the upper limit on the mass ($<~21.8 \rm~M_\oplus$); this yields a minimum TSM value of $>39$ (see supplementary material).
Of the now 15 known circumbinary exoplanets, TOI-1338/BEBOP-1b is the only one for which James Webb Space Telescope transmission spectroscopy can currently be pursued. 
%If we are to begin to unveil the mysteries of circumbinary exo-atmospheres, this particular Tatooine-like planet is currently our only hope.
If we are to unveil the mysteries of circumbinary Tatooine-like exo-atmospheres, the TOI-1338/BEBOP-1 system provides a new hope.

\begin{figure}
    \centering
	\includegraphics{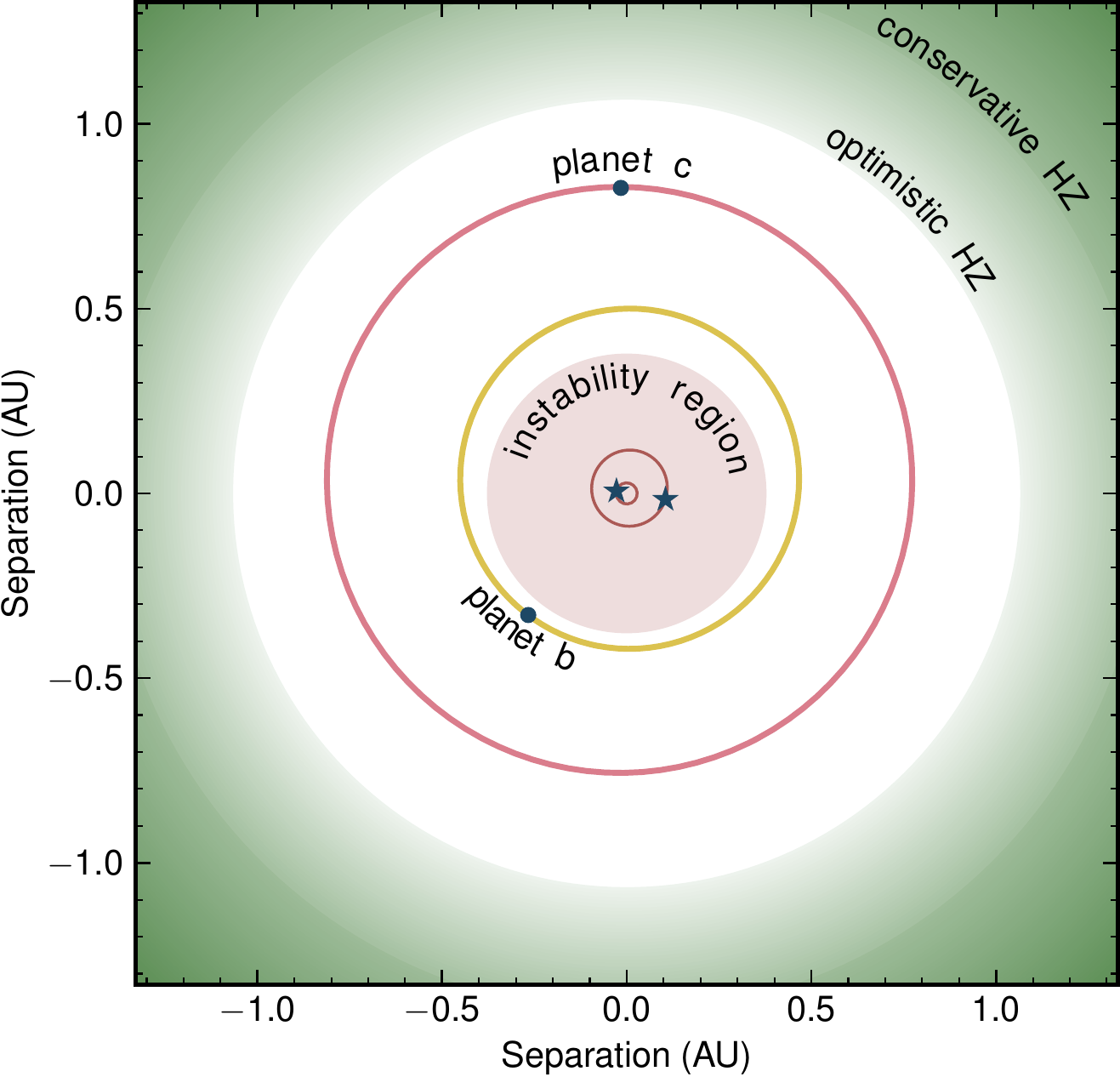}
    \caption{Overview of the TOI-1338/BEBOP-1 system along with the extent of the systems habitable zone calculated using the Multiple Star HZ website \cite{Muller2014}. The conservative habitable zone is shown by the dark green region, while the optimistic habitable zone is shown by the light green region.
    Binary stars are marked by the blue stars in the centre. TOI-1338/BEBOP-1c's orbit is shown by the red orbit models, based on $500$ randomly drawn posterior samples from a \texttt{kima} run, shaded from the $50^{\rm{th}}$ to $99^{\rm{th}}$ percentiles. TOI-1338/BEBOP-1b's orbit is shown by the yellow models, and is also based on $500$ random samples drawn from the posterior in its discovery paper \cite{Kostov2020}.}
    \label{fig:hab_zone}
\end{figure}

\begin{table}
	\centering
	\caption{TOI-1338/BEBOP-1 system orbital parameters from our analysis of the ESPRESSO and HARPS radial velocities, after removing outliers. $1\sigma$ uncertainties are provided as the last two significant digits, within brackets. Dates are given in BJD - 2,450,000. $^*$ from Kostov et al. (2020) \cite{Kostov2020}, $^{**}$ from Triaud et al. (2017) \cite{Triaud2017}.}
	\label{tab:results}
	\begin{tabular}{lll} 
		\hline
		\hline
		\multicolumn{3}{l}{\it Binary parameters}\\
		$P_{\rm bin}$        &day             & 14.6085579(57)\\
		$T_{0, \rm bin}$     &BJD             & 9287.20017(71)\\
		$K_{1, \rm bin}$     &$\rm km\,s^{-1}$ & 21.61764(73)\\
		$e_{\rm bin}$        &--              & 0.155522(29)\\
		$\omega_{\rm bin}$   &rad             & 2.05549(30)\\
		$\dot{\omega}_{\rm bin}$   &$\rm{arcsec\,year^{-1}}$ & 66.0$^{+65.0}_{-54.7}$\\
		$M_1$         &$\rm M_\odot$& 1.127(69)$^*$\\
		$R_1$         &$\rm R_\odot$& 1.345(46)$^*$\\
		$M_2$         &$\rm M_\odot$& 0.313(12)\\
% 		$R_2$         &$\rm R_\odot$& 0.3122(69)$^*$\\
		$a_{\rm bin}$ &$\rm AU$     & 0.1321(25)\\
		$T_{\rm eff, 1}$ &$\rm K$     & 6050(80)$^*$\\
		$V_{\rm mag, 1}$ &--     & 11.72(02)$^*$\\
		$\rm{Spectral type}$ &--     & F8$^{**}$\\
		\\
		\multicolumn{3}{l}{\it Planet b parameters}\\
		$P_{\rm b}$         &day             & 95.174(35)$^*$\\
% 		$T_{0, \rm b}$      &BJD             & --\\
		$K_{\rm b}$      &$\rm m\,s^{-1}$  & $<2.4(0.1)$\\
		$e_{\rm b}$         &--              & 0.0880(43) $^*$\\
% 		$\omega_{\rm b}$    &rad             & --\\
		$m_{\rm b}$  &$\rm M_{\oplus}$         & $<21.8(0.9)$\\
		$m_{\rm b}$  &$\rm M_{Jup}$         & $<0.0685(29)$\\
		$a_{\rm b}$ &$\rm AU$      & 0.4607(88)$^*$\\
		$\rho_{\rm b}$ &$\rm{g\,cm^{-3}}$&$<$0.36$^{+0.02}_{-0.01}$\\
		\\
		\multicolumn{3}{l}{\it Planet c parameters}\\
		$P_{\rm c}$         &day             & 215.5(3.3)\\
		$T_{0, \rm c}$      &BJD             & 9208.7(29.1)\\
		$K_{\rm c}$      &$\rm m\,s^{-1}$  & 5.6(1.0)\\
		$e_{\rm c}$         &--              & $<$0.16\\%for 50th percentile, <0.41 for 95%
		$\omega_{\rm c}$    &rad             & 3.84(86)\\
		$m_{\rm c} \sin i_{\rm c}$  &$\rm M_{\oplus}$& 65.2(11.8)\\
		$m_{\rm c} \sin i_{\rm c}$  &$\rm M_{Jup}$& 0.205(37)\\
		$a_{\rm c}$ &$\rm AU$      & 0.794(16)\\
		\hline
		\multicolumn{3}{l}{\it System parameters}\\
		$\gamma$             &$\rm km\,s^{-1}$  & 30.75335(99)\\
		\hline
		\multicolumn{3}{l}{\it Instrumental}\\
        $\rm{Jitter_{HAR}}$  &$\rm m\,s^{-1}$  & 4.2$^{+1.6}_{-3.0}$\\
        $\rm{Jitter_{ESP19}}$  &$\rm m\,s^{-1}$  & 4.2(2.0)\\
        $\rm{Jitter_{ESP21}}$  &$\rm m\,s^{-1}$  & 4.5(0.6)\\
        $\rm{RV\,Offset_{ESP19}}$  &$\rm m\,s^{-1}$  & -146.4$^{+1.9}_{-1.8}$\\
        $\rm{RV\,Offset_{ESP21}}$&$\rm m\,s^{-1}$  & -150.8(1.2)\\
		\hline
		\hline
	\end{tabular}
\end{table}

% {\it Science\/}'s publication workflow relies on Microsoft Word.  To
% translate \LaTeX\ files into Word, we use an intermediate MS-DOS
% routine \cite{tth} that converts the \TeX\ source into HTML\@.  The
% routine is generally robust, but it works best if the source document
% is clean \LaTeX\ without a significant freight of local macros or
% \texttt{.sty} files.  Use of the source file \texttt{scifile.tex} as a
% template, and calling {\it only\/} the \texttt{.sty} and \texttt{.bst}
% files specifically mentioned here, will generate a manuscript that
% should be eminently reviewable, and yet will allow your paper to
% proceed quickly into our production flow upon acceptance \cite{use2e}.

% Your references go at the end of the main text, and before the
% figures.  For this document we've used BibTeX, the .bib file
% scibib.bib, and the .bst file Science.bst.  The package scicite.sty
% was included to format the reference numbers according to *Science*
% style.

%BibTeX users: After compilation, comment out the following two lines and paste in
% the generated .bbl file. 

% \bibliography{scibib}

\section*{Author contributions}

The BEBOP project was established by DVM, and AHMJT, building on work by the WASP consortium that involved AHMJT, IB, ACC, MG, CH, PFLM, FP, AS and SU. The radial velocity (RV) data used in this manuscript were secured and obtained by MRS, AHMJT, LS, DVM, VK, DS, TAB, PFLM, NJM, and AS and their reduction involved MRS, AHMJT, FP and SU. The observational campaign was led by MRS for ESPRESSO observations, and AHMJT for HARPS. RV analysis was led by MRS with assistance from AHMJT, DVM, and JPF. Outliers in RV data were identified by TAB, and MRS. Detection limit analysis was performed by MRS, JH, and EL with help from JPF. TIP and FIP analysis was carried out by TAB, and NCH. Analysis of stellar activity was carried out by LS, MRS, and AHMJT. Independent stability analysis on the system was carried out by ACMC, and DVM with inputs from RM. Formation pathway simulations were carried out by GALC and RPN. The Transmission Spectroscopy Metric (TSM) was calculated by GD. The system was identified to have an inner transiting planet TOI-1338/BEBOP-1b by DVM, JAO and WFW. VK prepared the majority of figures in the paper. All co-authors assisted in writing and reviewing the manuscript. 

\section*{Acknowledgments}
We would like to thank the ESO staff at La Silla and Paranal for their continued support throughout this work, especially through the COVID pandemic. With special thanks to our ESPRESSO support astronomer M. Wittkowski. 

We also thank all of the observers who took part in the HARPS timeshare and were instrumental in collecting data for this projects. We particularly thank X. Dumusque and F. Bouchy for their work organising the timeshare.

This paper is based on observations collected at the European Southern Observatory under ESO programmes 103.2024, 106.216B, 1101.C-0721 and 106.212H. This research has made use of the services of the ESO Science Archive Facility. 

This work made use of the Astropy, numpy, pandas, scipy, corner, and matplotlib packages.

This research received funding from the European Research Council (ERC) under the European Union’s Horizon 2020 research and innovation programme (grant agreement n$^{\circ}$ 803193/BEBOP) and from the Leverhulme Trust (research project n$^{\circ}$ RPG-2018-418). 

MRS would like to acknowledge the support of the UK Science and Technology Facilities Council (STFC) (ST/T000295/1).

ACMC acknowledges support from
CFisUC (UIDB/04564/2020 and UIDP/04564/2020), GRAVITY (PTDC/FIS-AST/7002/2020),
PHOBOS (POCI-01-0145-FEDER-029932), and ENGAGE~SKA (POCI-01-0145-FEDER-022217),
funded by COMPETE 2020 and FCT, Portugal.
The stability maps were performed at the OBLIVION Supercomputer (HPC Center - University of \'Evora), funded by ENGAGE~SKA and by the BigData@UE project (ALT20-03-0246-FEDER-000033).

VK acknowledges support from NSF award AST2009501.

The project leading to this publication has received funding from the french government under the “France 2030” investment plan managed by the French National Research Agency (reference : ANR-16-CONV-000X / ANR-17-EURE-00XX) and from Excellence Initiative of Aix-Marseille University - A*MIDEX (reference AMX-21-IET-018). This work was supported by the "Programme National de Plan\'etologie" (PNP) of CNRS/INSU.

JPF is supported in the form of a work contract funded by national funds through Fundação para a Ciência e a Tecnologia (FCT) with reference DL57/2016/CP1364/CT0005.

ACC acknowledges support from STFC consolidated grant numbers ST/R000824/1 and ST/V000861/1, and UKSA grant number ST/R003203/1.

MG is FNRS Senior Research Associate

This research utilised Queen Mary's Apocrita HPC facility, supported by QMUL Research-IT (http://doi.org/10.5281/zenodo.438045). This work was performed using the DiRAC Data Intensive service at Leicester, operated by the University of Leicester IT Services, which forms part of the STFC DiRAC HPC Facility (www.dirac.ac.uk). The equipment was funded by BEIS capital funding via STFC capital grants ST/K000373/1 and ST/R002363/1 and STFC DiRAC Operations grant ST/R001014/1. DiRAC is part of the National e-Infrastructure.

%Here you should list the contents of your Supplementary Materials -- below is an example. 
%You should include a list of Supplementary figures, Tables, and any references that appear only in the SM. 
%Note that the reference numbering continues from the main text to the SM.
% In the example below, Refs. 4-10 were cited only in the SM.     
\section*{Supplementary materials}
Collected Observations\\
Modelisation of the data\\
Stellar Activity\\
System Dynamics\\
System Formation\\
Prospects for atmospheric follow-up with JWST\\
Journal Of Radial-Velocity Observations\\
Tables S1 to S4\\
Figures S5 to S16\\
References

\clearpage

% Include your paper's title here

\title{
\begin{center}
    \Huge{Supporting Online Material}
\end{center}}

% Place the author information here.  Please hand-code the contact
% information and notecalls; do *not* use \footnote commands.  Let the
% author contact information appear immediately below the author names
% as shown.  We would also prefer that you don't change the type-size
% settings shown here.

% \author
% {Matthew R. Standing,$^{1\ast}$ 
% Lalitha Sairam,$^{1}$ 
% Amaury H. M. J. Triaud,$^{1}$ %a.triaud@bham.ac.uk
% David V. Martin,$^{2,3}$\\ %martin.4096@osu.edu

% Thomas A. Baycroft,$^{1}$ %txb187@student.bham.ac.uk
% Gavin Coleman,$^{7}$ %gavin.coleman@qmul.ac.uk
% Alexandre C.~M. Correia,$^{5,6}$\\ %acor@uc.pt
% Georgina Dransfield,$^{1}$
% Vedad Kunovac,$^{4}$ %vedad.kunovac@lowell.edu
% Richard Nelson,$^{7}$ %r.p.nelson@qmul.ac.uk
% et al.,$^{8,9}$
% \\
% \normalsize{$^{1}$School of Physics and Astronomy, University of Birmingham, Edgbaston,}\\
% \normalsize{Birmingham B15 2TT, UK}\\
% \normalsize{$^{2}$Department of Astronomy, The Ohio State University, 4055 McPherson Laboratory,}\\
% \normalsize{Columbus, OH 43210, USA}\\
% \normalsize{$^{3}$NASA Sagan Fellow}\\
% \normalsize{$^{4}$Vedad affil}\\
% \normalsize{$^{5}$CFisUC, Departamento de F\'isica, Universidade de Coimbra, 3004-516 Coimbra, Portugal}\\
% \normalsize{$^{6}$IMCCE, UMR8028 CNRS, Observatoire de Paris, PSL Universit\'e,}\\
% \normalsize{77 av. Denfert-Rochereau,75014 Paris, France}\\
% \normalsize{$^{7}$Astronomy Unit, Queen Mary University of London, Mile End Road, London E1 4NS, UK}\\
% \\
% \normalsize{$^\ast$E-mail: mxs1263@bham.ac.uk}
% }
% Include the date command, but leave its argument blank.

\date{}

%%%%%%%%%%%%%%%%% END OF PREAMBLE %%%%%%%%%%%%%%%%

% Double-space the manuscript.

\baselineskip24pt

% Make the title.

\maketitle

% Place your abstract within the special {sciabstract} environment.

\begin{sciabstract}
    In this supplementary material we present additional methods and details in support of our main article ``Radial-velocity discovery of a second planet in the TOI-1338/BEBOP-1 circumbinary system". 
    The material is organised as follows, in Section~1 we describe the collected observations for the discovery, along with outlier removal. Section~2 presents the orbital fitting and data analysis techniques utilised, along with constraints on additional planetary companions in the system. Section~3 provides an overview of the stellar activity analysis on the system. In Section~4 we discuss the stability of the system, and in Section~5 we discuss formation pathways for the planetary system.  Section~6 hosts discussion on atmospheric follow-up observations of system. The Radial-Velocity data used in the analysis is available in Section~7.

\end{sciabstract}

% In setting up this template for *Science* papers, we've used both
% the \section* command and the \paragraph* command for topical
% divisions.  Which you use will of course depend on the type of paper
% you're writing.  Review Articles tend to have displayed headings, for
% which \section* is more appropriate; Research Articles, when they have
% formal topical divisions at all, tend to signal them with bold text
% that runs into the paragraph, for which \paragraph* is the right
% choice.  Either way, use the asterisk (*) modifier, as shown, to
% suppress numbering.

\section{Collected Observations}
TOI-1338/BEBOP-1 was selected for the Binaries Escorted By Orbiting Planets (BEBOP) programme \cite{Martin2019} from a large sample of low-mass eclipsing binaries identified by the EBLM project \cite{Triaud2017} under the name  EBLM~J0608-59. Those binaries were detected as part of the Wide Angle Search for Planets (WASP), as candidate transiting planet which were later shown with  using radial velocities to be ``false positive'' eclipsing binaries. To be part of the EBLM and BEBOP sample, a system needs to be a single-lined eclipsing binary. For identification and selection, the southern sample exclusively used the CORALIE spectrograph ($R\sim 45,000$; mounted on the 1.2m {\it Euler} Swiss telescope, at La Silla, Chile). TOI-1338/BEBOP-1 was first observed as a part of the EBLM project in 2009.

The BEBOP sample represents a sub-sample of the EBLM sample, where the binaries are selected to optimise planet-finding capability. The main discriminator is the obtained radial-velocity precision, which is typically a function of magnitude. There is also a bias to wider ($\gtrsim 5$~days) binaries, since rapid rotation due to tidal locking broadens spectral lines and reduces precision. Fortunately, the {\it Kepler} results in fact show a dearth of planets transiting the tightest binaries \cite{Martin2015}. We also progressively remove binaries where a long-term radial velocity trend reveals a presence of a third star, which is believed to have a detrimental impact on planet occurrence \cite{Martin2015}. Finally, systems are also removed based on simple activity indicators just as the line bisector \cite{Queloz2001}. Targets within the BEBOP sample receive both longer exposures than the rest of the EBLM sample, in order to improve precision. TOI-1338/BEBOP-1 was first observed as a part of the BEBOP project in 2014. Greater details about the selection process can be found in \cite{Martin2019}. 

There exists 55 radial velocities obtained with CORALIE, which are not used in this paper. Their typical uncertainty is $25~\rm m\,s^{-1}$, much higher than the signal we identify. Some of them were used to determine the spin--orbit angle of the binary \cite{Hodzic2020}.

\subsection{HARPS}
HARPS is a high-resolution, high-precision \'echelle spectrograph built for the detection of exoplanets \cite{Mayor2003}. It has a resolution $R\sim 100,000$ and typically achieves a long-term stability under $1~\rm m\,s^{-1}$. It is mounted on the ESO 3.6m telescope at La Silla, Chile.

The southern BEBOP sample, including TOI-1338/BEBOP-1 (under the name J0608-59) were observed by HARPS under two ESO large observing programmes (prog.ID 1101.C-0721 and 106.212H; PI Triaud). TOI-1338/BEBOP-1 received $61$ spectra between the dates of 2018-04-08 and 2022-04-18. Typically exposure times of 1800s were obtained, with a median radial-velocity precision of $5.73~\rm m\,s^{-1}$.

HARPS data were reduced by HARPS Data Reduction Software (DRS) version 3.5 (which is hosted at the Observatory of Geneva, and will be made public in a few months). A description of how the DRS works can be found in \cite{Baranne1996} and in \cite{Standing2022}. Spectra are correlated using a weighted numerical mask matching the spectral type of the target producing a cross-correlation function (CCF). A Gaussian function is fitted to the CCF to find the mean radial velocity. 

Version 3.5 of the DRS is not very different from the version held by ESO but it allowed us to recorrelate the spectra with our own specification. The standard DRS is built to study single, slow rotating stars. It assumes two quantities: a common mean radial velocity for all spectra of a given system, and a correlation window of $30~\rm km\,s^{-1}$ on either side of that mean velocity. In the case of binaries, the velocity changes according to the binary phase. In addition some of the BEBOP sample targets rotate fast enough that a $30~\rm\,km\,s^{-1}$ window is not adequate. Using version 3.5 allowed us to centre the correlation window to the observed velocity at each epoch, thus ensuring the same set of absorption lines are used to produce the radial velocities, epoch after epoch. For TOI-1338/BEBOP-1 we used the standard $30~\rm km\,s^{-1}$ correlation window and a G2 mask to produce the cross-correlation window. The reduction software provides the radial velocity, its uncertainty (determined from photon noise and the line width), the line width (FWHM), the span of the bisector slope (Bis\_Span), and automatically produced corrections to the barycentre of the Solar system.

All HARPS data are available at the ESO public archive by searching for J0608-59, and can be found in Table~3 in this supplementary material.

\subsection{ESPRESSO}
Following the discovery of TOI-1338b \cite{Kostov2020}, $20$ radial-velocity measurements were obtained in 2019 with the ESPRESSO spectrograph located in Paranal, Chile \cite{Pepe2021} in an attempt to confirm the planet and more accurately constrain its mass (Prog.ID 103.2024, PI Triaud). Exposure times were typically $900~\rm s$ in length and yielded a median precision of $2.83~\rm m\,s^{-1}$. These data combined with previously obtained HARPS spectra yielded no detection of TOI-1338/BEBOP-1b, but %though allowing a $3\sigma$ upper limit on the mass to be determined as $25~\rm{M_{\oplus}}$ and hinting 
hinted at the presence of an outer companion planet with an orbital period of $\approx200$~days. 
To confirm this candidate planet and attempt to detect the inner transiting planet, a further $103$ ESPRESSO spectra were obtained with exposure times of $900~\rm s$ yielding a median precision of $2.63~\rm m\,s^{-1}$ (Prog.ID 106.216B PI Standing). Of these $103$ measurements, $3$ were obtained during the primary eclipse of the binary, and $1$ during the transit of planet b. These $4$ observations have been discarded from our analysis as they are affected by the Rossiter-McLaughlin effect, which we do not model here \cite{Triaud2018, Hodzic2021}.

In our radial-velocity analysis, we fit for an offset between ESPRESSO data obtained in $2019$, and those in $2021$, post COVID closures. This $4.6~\rm m\,s^{-1}$ offset, though small, was introduced by calibration lamp change during the COVID closure of telescope in $2020$ \cite{Faria2022}. Our $2019$ dataset was obtained after the fibre change on the instrument in $2019$. Without accounting for this offset the planetary signal is still detected in the data.
COVID closure of both the La Silla and Paranal observatories delayed the collection of data on this system and by consequence the discovery of this planet by a year. Complete phase coverage of TOI-1338/BEBOP-1c's orbit was finally achieved when the final data points were collected in March $2022$.% under Prog.ID 106.216B.003 PI Standing.

ESPRESSO data were reduced using version 2.3.3 of the ESPRESSO pipeline (publicly available at ESO) with a procedure similar to that used to HARPS, but adapted to ESPRESSO. Each cross correlation function was obtained with a G2 template spectra, step size of $0.5~\rm{km\,s^{-1}}$, and a $300~\rm{km\,s^{-1}}$ window centered on the radial velocity of the binary.
To summarise, we have $123$ ESPRESSO observations in total, with $20$ observations in ESPRESSO $2019$, and $103$ in ESPRESSO $2021/2022$. The total timespan of the HARPS and ESPRESSO combined is $1472$~days. 

All ESPRESSO data are available at the ESO public archive by searching for J0608-59. All radial-velocity data used in our analysis can be found in Tables~2-4 in this supplementary material.

\subsection{TESS}

TOI-1338/BEBOP-1 was a target for the radial-velocity survey before the launch of TESS. It was then later discovered to contain a 95-day transiting planet (TOI-1338b, aka TOI-1338/BEBOP-1b), based on four sectors of TESS data (three in short cadence, 120 seconds). More data has been taken in the years since. These data are mentioned here for completeness but we do not include them in the analysis; TOI-1338/BEBOP-1c is an independent radial-velocity discovery.

\subsection{\textbf{Outlier treatment}}
Outliers were identified using two methods. The first involved a-priori finding outliers in the span of the bisector slope (Bis\_Span, a measure of line shape \cite{Queloz2001}) and Full Width at Half Maximum (FWHM), and excluding these data. The radial-velocity analysis in this paper used this method. The second method used a student-t distribution to account for outliers as part of the sampling process (e.g. \cite{Agol2021}), we used this to check the results from the first method.

Preliminary outliers were first removed where the wrong star was observed or the observations occured during an eclipse or a planetary transit. The following methods to deal with outliers were then applied to the remainder of the data.

For the a-priori removal, each set of points was fit as a mixture model with inlier and outlier populations, following the method described in Hogg et al. (2010) \cite{Hogg2010}. For both FWHM and Bis\_Span, the inlier model was simply a constant with a small scatter allowed in addition to the uncertainties (which were taken as double the radial-velocity uncertainty). The outlier model also fits a constant but with a very wide population scatter.

We used PyMC3 \cite{Salvatier2015} to fit a mean and scatter for both the inlier and outlier distributions, as well as a parameter \(f\) for the proportion of points that are outliers, the prior on \(f\) is a \(\beta[1.5,9]\) distribution favouring a low proportion of outliers. These parameters were then used to calculate the probability of the \(i^{th}\) point being an outlier for each posterior using the following equation:
\begin{equation}
    P_{i,out} = \frac{f L_{i,out}}{f L_{i,out}+(1-f)L_{i,in}}
\end{equation}
where \(L_{i,out}\) and \(L_{i,in}\) are the likelihoods for the \(i^{th}\) point in the respective outlier and inlier distributions. 
% An example of this is shown in Figure~\textbf{XX} for the HARPS data.

% \begin{figure}
%     \centering
%     \includegraphics[width=\columnwidth]{Outliers.png}
%     \caption{Bottom: Bis\_Span for the Harps data. Top: Corresponding outlier probability. Inliers are blue and the only outlier is in red}
%     \label{fig:my_label}
% \end{figure}

The power of this method is in removing the human factor from outlier identification, and not having to resort to sigma-clipping which does not usually consider measurement uncertainties. A total of 17 outliers were identified this way.

For the student-t method, we ran the fits on the dataset accounting for outliers using a student-t distribution. 
%This distribution places less emphasis on strong outliers, and as such these outliers do not affect the results as much as if a Gaussian distribution was being used. 
Outliers in radial velocity can then be directly identified, rather than relying on the Bis\_Span and FWHM indicators as in method 1. This student-t method identified 6 outliers of which 3 are shared with the 17 outliers identified in method 1; the 3 points are particularly strong outliers in the student-t analysis.

Importantly, the resulting detection of TOI-1338/BEBOP-1c and non-detection of TOI-1338/BEBOP-1b is not affected by the method of dealing with outliers, and all the parameters from the student-t fit were consistent with those found removing the outliers a-priori.

In summary: 3 outliers were identified in the HARPS data in FWHM one of which is also identified in Bis\_Span and is such a strong outlier it is suspected that the wrong star was observed; 1 outlier was identified in the ESPRESSO $2019$ data in FWHM; and 13 outliers are identified in the ESPRESSO $2021/2022$ data in FWHM (2 of which are also identified in Bis\_Span). Data points identified as outliers are flagged in Tables~2-4 of this supplementary material. We denote outliers, excluded from the analysis, with the following flags: Wrong star W, Bis\_Span B, FWHM F, During binary transit R, during planetary transit P.

\section{Modeling of the data}
\subsection{Kima}

For the radial-velocity analysis we use the \texttt{kima} package \cite{Faria2018}. \texttt{kima} models radial-velocity data with a sum of Keplerian functions from $N_{\rm{p}}$ orbiting planets, and estimates the posterior distributions for each of the orbital parameters.

To sample the posterior distribution \texttt{kima} uses a Diffusive Nested Sampling (DNS) approach \cite{Brewer2011}. This provides \texttt{kima} with estimates on the evidence for each model, allowing for model comparison \cite{Feroz2011, Brewer2014}.
This model comparison can then be used to compare the ratio of evidences between models with different numbers of Keplerian signals. Importantly, the number of planets $N_{\rm{p}}$ is a free parameter like any other.

The radial-velocity analysis follows that described in Standing et al. (2022) \cite{Standing2022} and Triaud et al. (2022) \cite{Triaud2022} with added post-Keplerian corrections as described in Baycroft et al. (in prep). These corrections account for relativistic and tidal effects on the radial velocity measurements from the binary orbit which have been derived and tested in \cite{Zucker2007, Konacki2010, Arras2012, Sybilski2013, Standing2022}. In addition, we directly fit for apsidal precession of the binary within \texttt{kima}, which is the largest Newtonian effect we can expect. We find a value of $\dot{\omega}_{\rm bin} = 66.0^{+65.0}_{-54.7}~\rm{arcsec\,year^{-1}}$, distinct but barely different from zero. Details about this implementation are described in Baycroft et al. (in prep). Prior distributions used in our analysis can be found in Table~\textbf{2}.

\begin{table}
	\centering
	\caption{Prior distributions for the binary and planetary RV model signals in \texttt{kima}}
	\begin{tabular}{c c c c}
		\hline
		\hline
	    Parameter & Unit & \multicolumn{2}{c}{Prior distribution}\\
		\hline
		    &   & Binary & Planet\\
		$N_{\rm p}$ &  & $1$ & $\pazocal{U}(0,3)$\\
		$P$ & days & $\pazocal{U}(14.61\pm{0.01})$ & $\pazocal{LU}(4\times P_{\rm bin},2000)$\\
		$K$ & $\rm{m\,s^{-1}}$ & $\pazocal{U}(21617\pm{15})$ & $\pazocal{MLU}(0.1, 100)$\\
		$e$ & & $\pazocal{U}(0.1556\pm{0.001})$ & $\pazocal{K}(0.867, 3.03)$\\
		$\phi$ & & \multicolumn{2}{c}{$\pazocal{U}(0,2\pi)$}\\
		$\omega$ & & \multicolumn{2}{c}{$\pazocal{U}(0,2\pi)$}\\
		$\dot{\omega}_{\rm bin}$ & $\rm{arcsec\,year^{-1}}$ & $\pazocal{N}(0,1000)$ & - \\
		$\sigma_{\rm jit}$ & $\rm{m\,s^{-1}}$ & \multicolumn{2}{c}{$\pazocal{LU}(0.01,67)$}\\
		$\gamma$ & $\rm{m\,s^{-1}}$ & \multicolumn{2}{c}{$\pazocal{U}(30761\pm{100})$}\\
		\hline
	\end{tabular}
	\label{tab:prior distributions}
%    \begin{tablenotes}
    \small
    \begin{description}
        \item \textbf{Notes:} $N_{\rm p}$ denotes the Number of Planetary Keplerian signals to fit to the data. $P_{\rm bin}$ denotes the Period of the binary. $\pazocal{U}$ denotes a uniform prior with an upper and lower limit, $\pazocal{LU}$ is a log-uniform (Jeffreys) prior with upper and lower limits, $\pazocal{MLU}$ is a modified log-uniform prior with a knee and upper limit, $\pazocal{N}$ is a gaussian prior with a mean and standard deviation, and $\pazocal{K}$ is a Kumaraswamy prior \cite{Kumaraswamy1980} which takes two shape parameters \cite{Kipping2013}.
    \end{description}
%    \end{tablenotes}
\end{table}

\subsection{Resulting fit}

The \texttt{kima} combined fit of our HARPS and ESPRESSO data favours a single planet model in addition to the binary orbit with a Bayes factor of $>29,000$.
The Bayes factor is the ratio of the Bayesian evidence between two competing models. In our case the Bayes factor is the number of posterior samples obtained with $\rm{N_p}$ planetary signals, over those with $\rm{N_p}-1$. The Bayes factor value indicates the measure of support in favour of one model over the other \cite{Kass1995,Trotta2008,Standing2022}.

A Bayes factor value of $>150$ indicates very strong evidence \cite{Jeffreys1961,Standing2022} in favour of a single planet in addition to the binary orbit.

When searching for the inner transiting planet in the data we obtain a Bayes factor of $1.4$ in favour of two planetary signals. This is categorised as inconclusive evidence of any further signals in the data. 

Forcing a fit on a $95$~day Keplerian signal yields no clear signal corresponding to TOI-1338/BEBOP-1b with posterior samples having a semi-amplitude consistent with $0~\rm{m.s^{-1}}$. This fit still produces a clear detection of TOI-1338/BEBOP-1c.

To obtain orbital parameters for the system, we follow the same procedure as detailed in Standing et al. (2022) \cite{Standing2022}. Proposed posterior samples with orbits which cross one another, or into the instability region of the binary, are removed. The remaining samples are clustered with the \texttt{HDBSCAN} clustering algorithm \cite{McInnes2017}. Clusters corresponding to the binary and TOI-1338/BEBOP-1c are then plotted using the \texttt{Corner} package \cite{corner}. Orbital parameters are then determined as the $50^{\rm th}$ percentile, with $1\sigma$ uncertainties estimated from the $14^{\rm th}$ and $84^{\rm th}$ percentiles of the cluster. Corner plots for the binary and TOI-1338/BEBOP-1c can be seen in Figures~\textbf{5} and \textbf{6} respectively.

\subsection{Constraints on TOI-1338/BEBOP-1b and further planetary companions}
Using \texttt{kima} as in Standing et al. (2022), and Triaud et al. (2022) \cite{Standing2022, Triaud2022} we calculate a detection limit on additional undetected planetary signals in the radial-velocity data. 
Firstly, we remove the highest likelihood Keplerian model corresponding to TOI-1338/BEBOP-1c from the data. Following this, to calculate the detection limit we fix $N_{\rm{p}}=1$ in our fit, include the binary with tight priors on its parameters, and obtain posterior samples of all remaining signals which are compatible with the data. 
This method is conceptually similar to that described in Tuomi et al. (2014) \cite{Tuomi2014}, where the sampler is forced to fit an additional signal to the data. Any signals found are compatible with the data, but as of yet undiscovered, since only one planetary signal is present in our original fit.

The resulting posterior samples can be seen as a greyscale density plot in Figure~7, with a blue $99\%$ contour corresponding to our detection limit. The blue limit demonstrates that we are sensitive to additional sub-Saturn mass planets for periods out to $2000$~days, while we are sensitive to Neptune mass planets near the instability limit.

We note a density of posterior samples at periods around $100$~days in agreement with the presence of TOI-1338/BEBOP-1b seen in \textit{TESS} photometry \cite{Kostov2020}. 
Following the findings described in Standing et al. (2022) \cite{Standing2022}, we calculate an additional detection limit to place an upper limit on the mass of TOI-1338/BEBOP-1b. This limit is calculated in the same manner as the blue line, though only using posterior samples with eccentricities $<0.1$ (the approximate upper limit on the eccentricity of TOI-1338/BEBOP-1b \cite{Kostov2020}, and a value allowed by our orbital stability analysis). A running mean is applied on the red eccentricity cut line to ensure $>1000$ samples are in each bin. 
The resulting red detection limit demonstrates our sensitivity to circular planets at the orbital period of TOI-1338/BEBOP-1b is $\approx2.4\pm0.1~\rm{m.s^{-1}}$. This allows us to place an upper limit on the mass of TOI-1338/BEBOP-1b of $21.8\pm0.9~M_{\oplus}$. With a radius of $\approx6.9~R_{\oplus}$, we calculate the mean density of TOI-1338/BEBOP-1b to be $<0.36~\rm{g.cm^{-3}}$.
The mass of planet b from Kostov et al. \cite{Kostov2020} is primarily determined thanks to the apsidal precession imprinted on the eclipse timing variations, under the assumption of a single planet. Our study uses exclusively radial velocities. Since they are not sensitive to the planet’s mass, we only place an upper limit.

The uncertainty of the detection limit is determined by performing numerical experiments with {\tt kima}. We first generate a radial-velocity timeseries, following a Gaussian distribution, with dates following a log-uniform distribution. Then we run {\tt kima} just like any other system and produce a large number of posteriors (in our case 685,000). Following that, we  compute a detection limit like in Standing et al. (2022) \cite{Standing2022} and call this our fiducial case. The next step is to calculate detection limits for many subsamples of the posterior and measure their fractional distance to the fiducial detection limit. We find that the fractional error in the position of subsamples' detection limits follows a square root law. Consequently, we fit these data with a function $y = a\,x^{-1/2}$, where $y$ is the fractional uncertainty of the detection limit at a given orbital period and $x$ is the number of posterior within a given posterior subsamples. We find $a = 15.1\pm0.3$. We use this relation to represent an uncertainty on Fig. 4.

\begin{figure}
    \centering
    \includegraphics[width=\columnwidth]{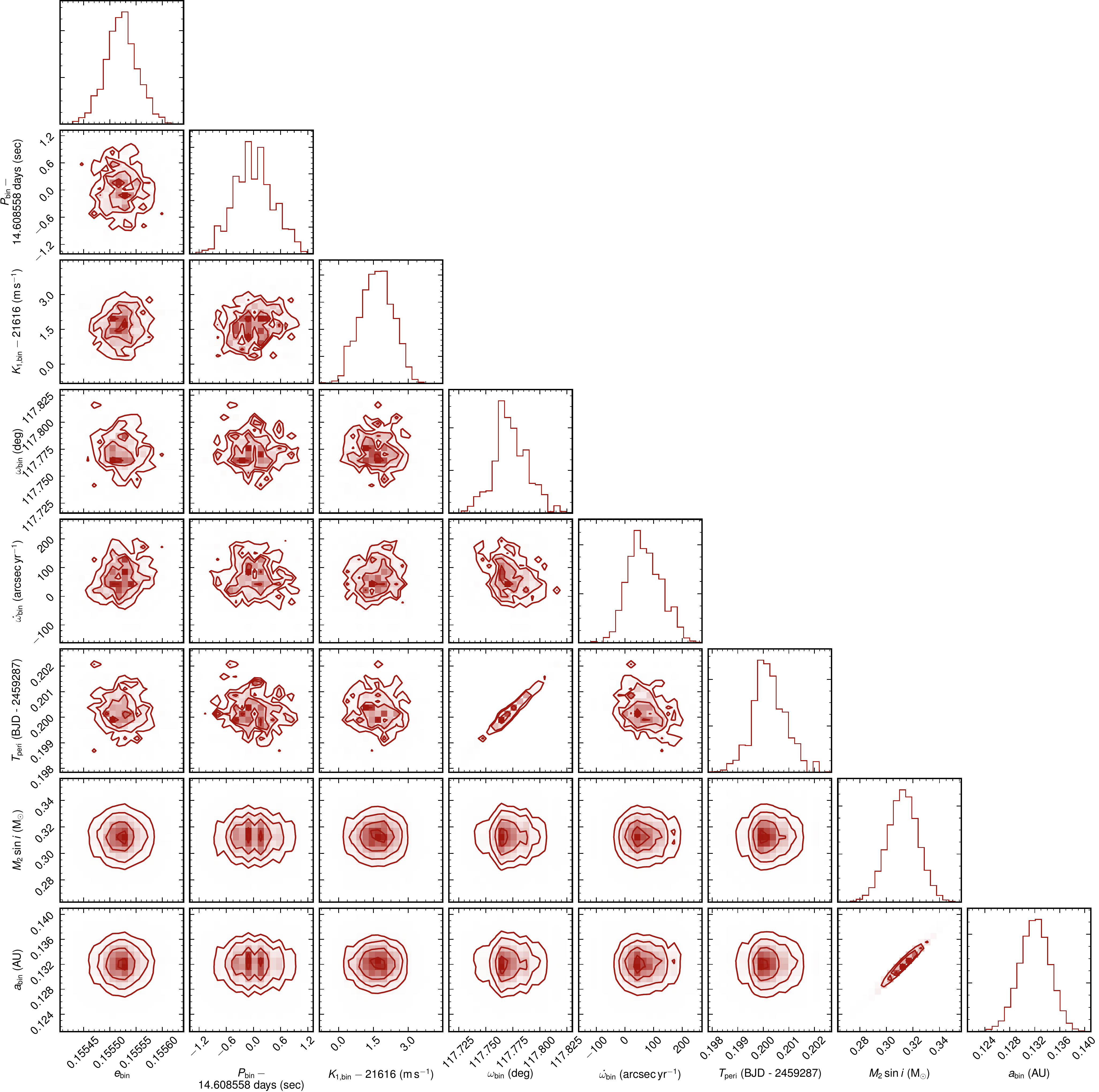}
    \caption{Corner plot of posterior sample distributions for the Binary stars orbital parameters; eccentricity, orbital period, semi-amplitude, argument of periastron, binary apsidal precession, time of periastron passage, $M\sin i$ in Solar masses, and semi-major axis.}
\end{figure}

\begin{figure}
    \centering
    \includegraphics[width=\columnwidth]{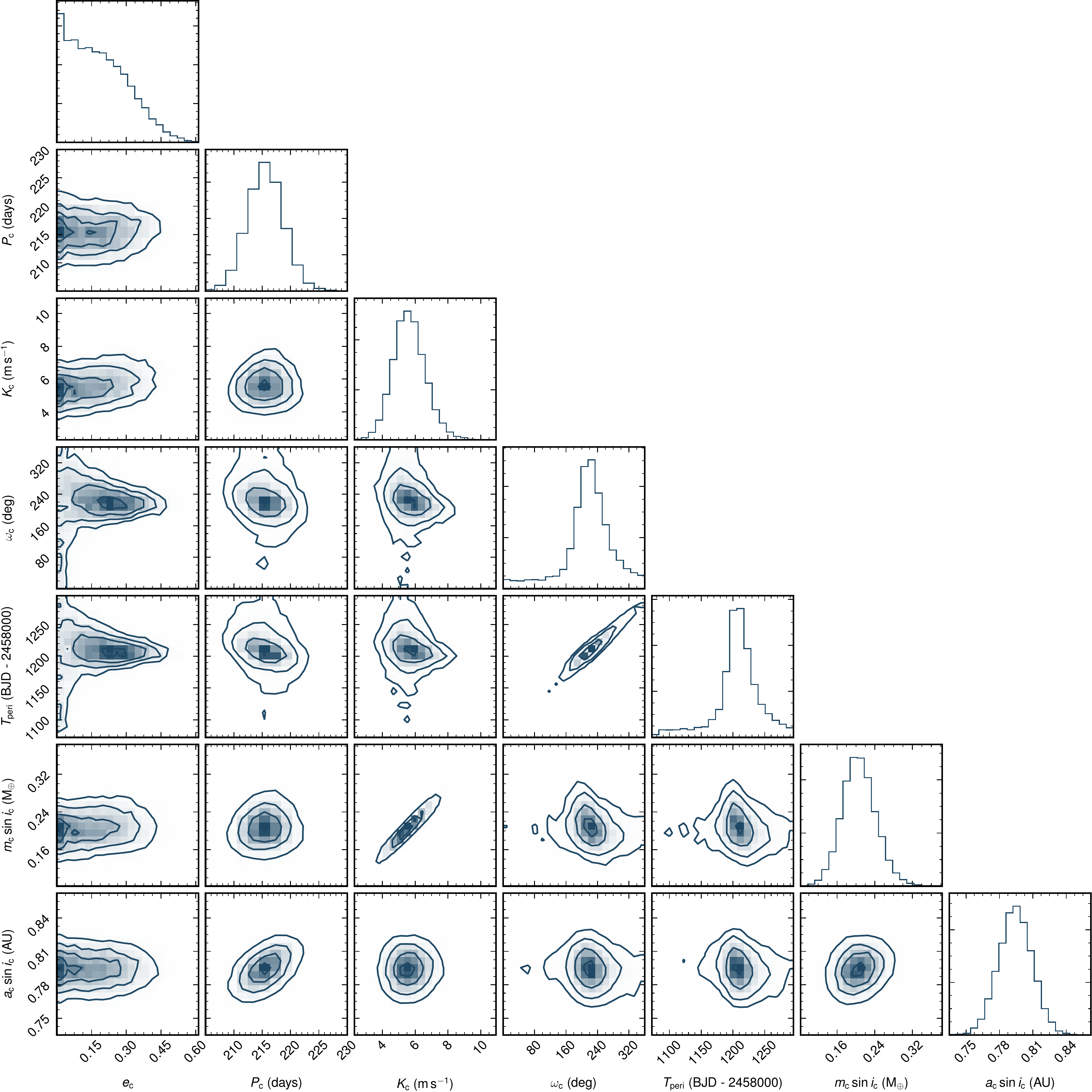}
    \caption{Corner plot of posterior sample distribution for TOI-1338/BEBOP-1c orbital parameters; eccentricity, orbital period, semi-amplitude, argument of periastron, time of periastron passage, $M\sin i$ in Earth masses, and semi-major axis.}
\end{figure}

\begin{figure}
    \centering
    \includegraphics[width=\columnwidth]{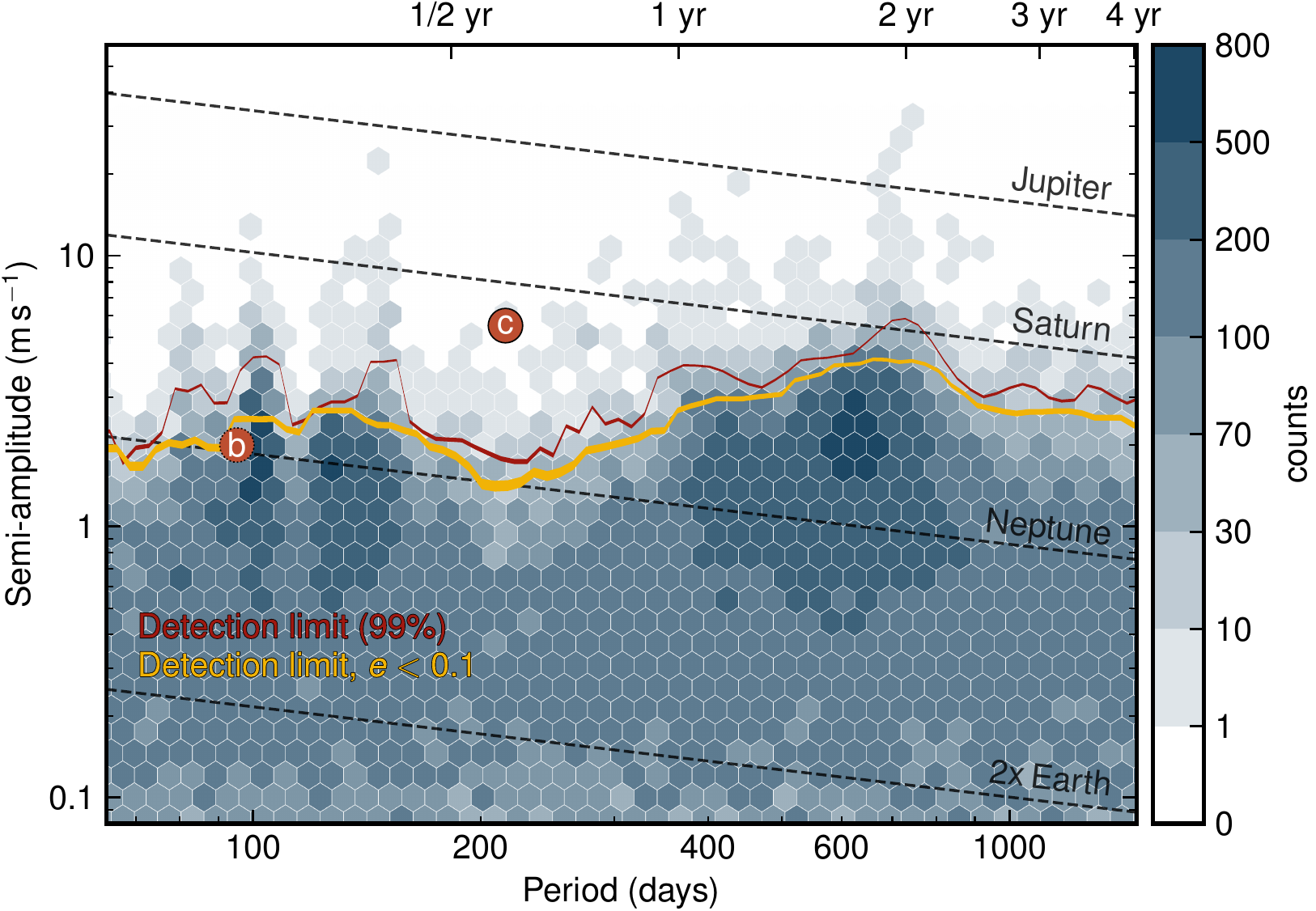}
    \caption{Hexbin plot denoting the density of $\sim100\,000$ posterior samples obtained from four separate \texttt{kima} runs on TOI-1338/BEBOP-1 Radial-Velocity (RV) data with $N_{\rm{p}}$ fixed $=1$. The red line shows the calculated $99\%$ detection limit, along with its associated error. Point `c' indicates the position of detected planet TOI-1338/BEBOP-1c, and point `b' that of planet TOI-1338/BEBOP-1b undetected in RV data alone. The orange line shows the detection limit on samples with eccentricity $<0.1$ (the upper limit on TOI-1338/BEBOP-1b's eccentricity \cite{Kostov2020}).}\label{detectionlimit}
\end{figure}

\subsubsection{FIP and TIP}

We compute the True Inclusion Probability (TIP) and False Inclusion Probability (FIP) for a signal being present in the data over various frequency bins \cite{Hara2021}. Plotting this in a periodogram gives similar information to the detection limit described above. However, where the detection limits give us information about an upper limit on the mass of a potential planetary signal, the TIP periodogram effectively gives us the probability of there being a planetary signal at that given period. It has been shown that using this metric as a detection criterion is optimal in the context of exoplanet detection \cite{Hara2022}.
We find 1 peak in the TIP periodogram further to the detected planet c at: \(\approx700\) days with TIP of 0.94.
This is shown in Figure~8 which shows the FIP and TIP periodograms. We attribute this signal to the window function as seen in Figure~9, though further observations will provide greater insight into the source of the signal.

\begin{figure}
    \centering
    \includegraphics[width=\columnwidth]{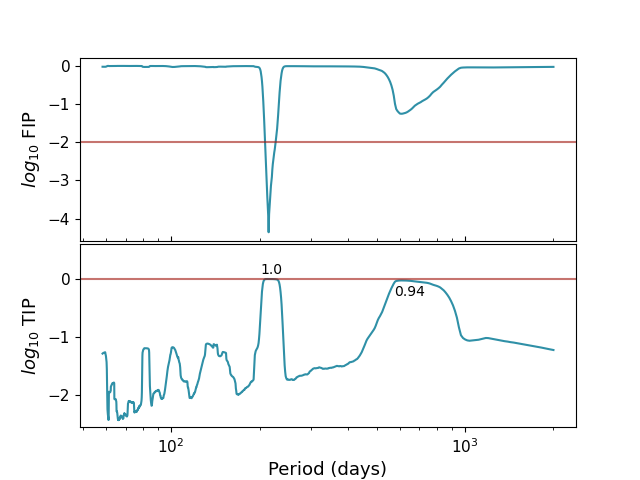}
    \caption{Top: False Inclusion Probability (FIP) periodogram showing the clear detection of TOI-1338/BEBOP-1c at $\approx215$~days. Bottom: True Inclusion Probability (TIP) periodogram showing the significance of any other peaks with TIP \(>0.2\). The second highest peak occurs at $\approx700$~days with a significance of 94\%.}
\end{figure}

\section{Stellar Activity}

Chromospheric emission lines observed in the stellar spectrum often trace stellar magnetic activity. Here we describe the activity indices measured and used in our analysis. We used the open-source package {\tt actin} \cite{Gomes2018} to measure the Ca II H\&K, He I, sodium D doublet (Na D a and b), Ca I and H$\alpha$ activity indices. 

First we compute a generalised Lomb-Scargle (GLS) periodogram \cite{Zechmeister2009}
of each activity indicators, from individual instruments, and using the combined dataset. 
In Figure~10 (a,b), we show the window function of each individual and combined datasets and the periodogram of radial velocity. 
We compute false alarm probability levels of   $10\%$, $1\%$ and $0.1\%$ using a bootstrap randomisation of the dataset. We consider a signal to be significant if the false alarm probability level is $<0.1\%$. We searched for periodic signals of chromospheric activity indicators to investigate the possibility of TOI-1338/BEBOP-1c's period $P_{\mathrm{c}}$ being produced by magnetic activity of TOI-1338/BEBOP-1A (the primary). 
Figure~10 (c-g) depicts the periodogram of Ca II, H$\alpha$, Na D, He I and Ca I indices. None of the activity indicators shows any significant signal at or near P$_{\mathrm{c}}$ ($\sim$ 215 d). We see significant period at $\sim$ 7.5 d in Na D (FAP $\sim$0.1\%) and He I (FAP $\sim$1\%), however, the nature of these signals are not obvious. We see weak signals in H$\alpha$ at 270~d caused by a combination of window function and a long-period term, however, the signal is not significant. In addition, the activity time series shown in Figure~11 does not show any obvious modulation. 

We searched All-Sky Automated Survey (ASAS; \cite{Pojmanski_1997}) archival data to constraint the rotation period and to investigate the nature of the TOI-1338/BEBOP-1c signal. The ASAS observations spanning over ~10 years. The ASAS data shows a strong signal at ~61 d (10\% false alarm probability) which could be related to stellar rotation ($\sim$3P$_{\mathrm{rot}}$). 

In addition, analysis of the radial-velocity bisector shows no correlation with the $\sim215$~d signal, yielding a Pearson Correlation co-efficient of $<0.052$.

The ASAS data and activity indicators do not show any significant signal at $\sim215$~d, therefore this signal is likely planetary in nature.

\begin{figure}
    \centering
	\includegraphics[width=0.8\columnwidth]{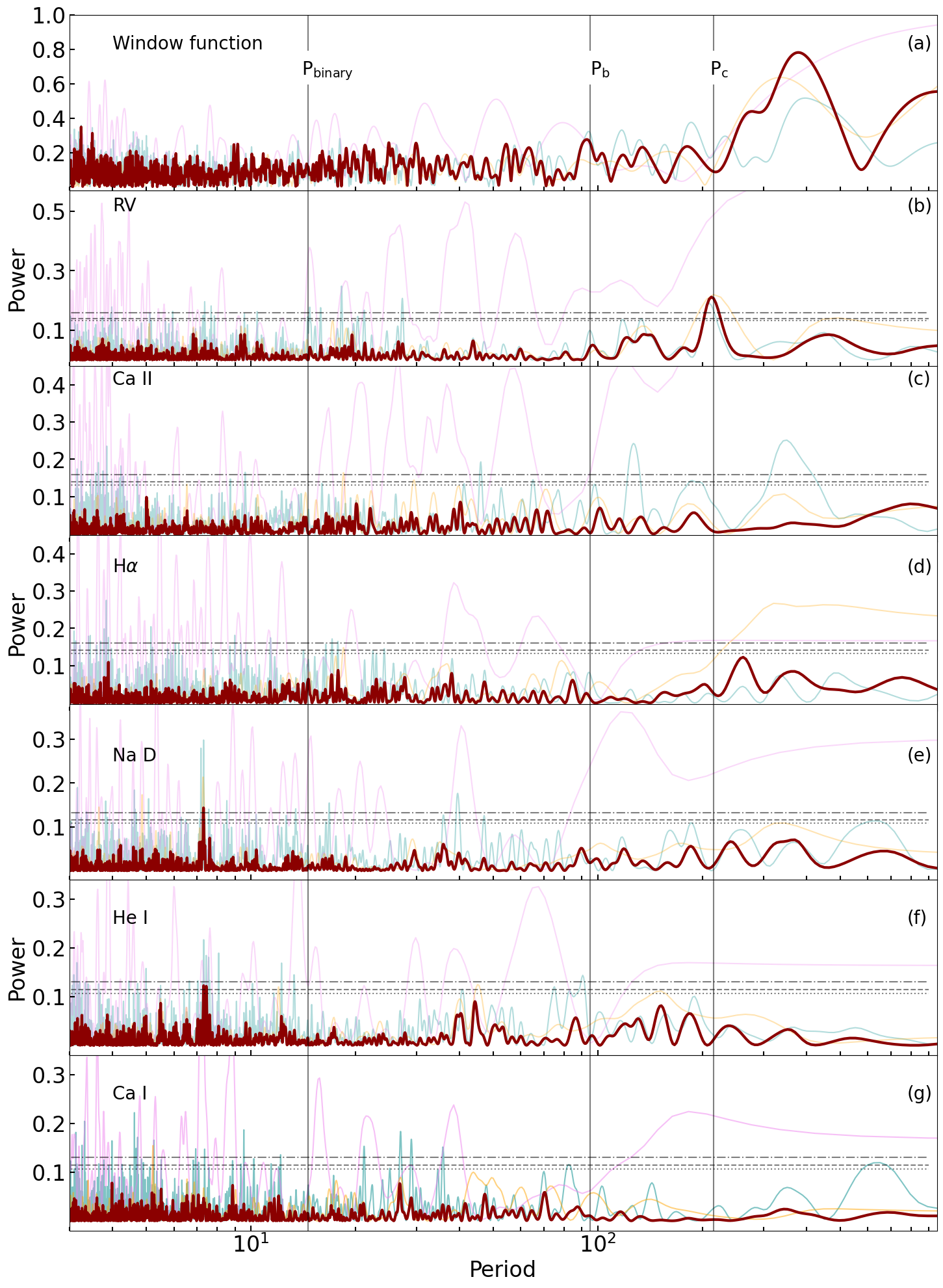}
    \caption{Generalised Lomb-Scargle periodogram of TOI-1338/BEBOP-1 spectroscopic data. The top panel shows the window function of the combined dataset (dark red). 
    b, Periodogram of the radial-velocity measurements. c-g, Periodogram for Ca II, H$\alpha$, Na D, He I and Ca I index. 
    The cyan, violet and orange curves corresponds to HARPS, ESPRESSO 2019 and ESPRESSO 2021/2022, respectively, and the dark red curves correspond to the combined dataset. The horizontal lines represent the bootstrapped false alarm probability levels of 10\% (dotted), 1\% (dashed), and 0.1\% (dotted dash). The vertical solid lines indicate the binary period and planetary periods  (P$_{\mathrm{b}}$ and P$_{\mathrm{c}}$).}
    \label{fig:periodogram}
\end{figure}

\begin{figure}
    \centering
	\includegraphics[width=0.8\columnwidth]{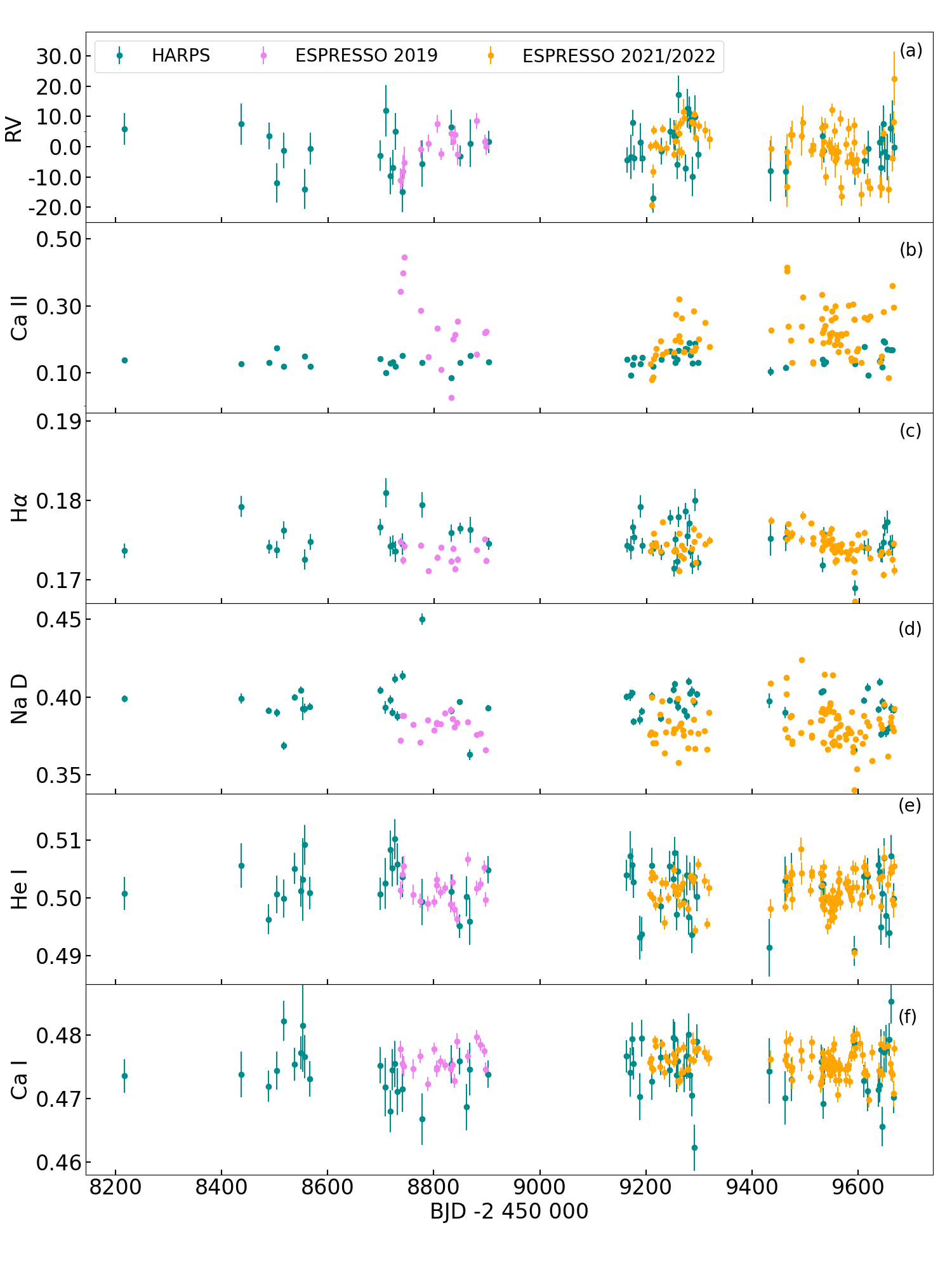}
    \caption{Radial velocity, Ca II, H$\alpha$, Na D, He I and Ca I timeseries for TOI-1338/BEBOP-1.}
    \label{fig:activity_ts}
\end{figure}

\section{System dynamics}

In this section we deal with various aspects of the orbital dynamics of TOI-1338/BEBOP-1, first looking at the stability of the system, and then estimating whether the outer planet, TOI-1338/BEBOP-1c, will show transits at some point into the future.

\subsection{Orbital Stability}\label{subsec:discussion_stability}
%%%%%%%%%%% Alex / David %%%%%%%%%%% 

The orbital solution given in Table~1 in the main text shows a compact two-planet circumbinary system ($a_\mathrm{bin} = 0.13$, $a_\mathrm{b} = 0.46 $ and $a_\mathrm{c} = 0.79$~au). 
The inner planet has a maximum mass similar to Neptune ($m_\mathrm{b} = 21.8$~M$_\oplus$) and is close to the circumbinary stability boundary \cite{Quarles2018, Kostov2020}, while the outer planet minimum mass is similar to Saturn  ($m_\mathrm{c} = 0.21$~M$_\mathrm{Jup}$).
As a consequence, we expect strong mutual gravitational interactions between the stars and the planets.
In order to study the stability of the system, we performed a global frequency analysis \cite{Laskar1990, Laskar1993} in the same way as achieved for other circumbinary planetary systems \cite{Correia2005}.
In our analysis, we always consider for the inner planet its maximal mass, which can be seen as a superior limit for stability.
Moreover, the mass of this planet is relatively small and does not impact much the stability, and so smaller masses do not change much the global picture described here.

The system is integrated on a regular 2D mesh of initial conditions in the vicinity of the best fit (Table~1 main text).
We used the symplectic integrator SABAC4 \cite{LaskarRobutel2001}, with a step size of $0.001$~yr and general relativity corrections.
Each initial condition is integrated for 5000~yr, and a stability indicator, $\Delta = |1-n_{b'}/n_b|$, is computed. Here, $n_b$ and $n_b'$ are the main frequency of the mean longitude of the inner planet over 2500~yr and 5000~yr, respectively, calculated via the frequency analysis \cite{Laskar1993}.  Note that while the osculating mean motion varies over a single planetary orbit as a result of energy exchange with the binary and planet c, $n_b$ calculated over time intervals much longer than any resonant variations is constant for stable systems, while it drifts for unstable systems until one of the planets escapes. The results are reported in color, where yellow represents strongly chaotic trajectories with $\Delta > 10^{-2}$, while extremely stable systems for which $\Delta < 10^{-8}$ are shown in purple/black. Orange indicates the transition between the two, with $\Delta \sim 10^{-4}$.

% to be the variation in the measured mean motion of the inner planet over the two consecutive 2500~yr intervals of time \cite{Couetdic2010}.
% For regular motion, there is no significant variation in the mean motion along the trajectory, while it can vary significantly for chaotic trajectories.

Observationally, only the inner planet's eccentricity is well-constrained: $e_\mathrm{b} = 0.088 \pm 0.004$ \cite{Kostov2020}. In contrast, only an upper limit can be placed for the outer, non-transiting planet: $e_\mathrm{c} < 0.1$ (Table~1 main text).

Therefore, in a first experiment, we explore the stability of the system in the plane $(e_\mathrm{b}, e_\mathrm{c})$, assuming coplanar orbits.
The results are shown in Figure~11.
We observe that that the eccentricities of both planets must be smaller than $0.1$ to ensure a stable system.
Moreover, since we already have for the inner planet $e_\mathrm{b} \approx 0.1$ \cite{Kostov2020}, the only remaining possibility for the outer planet is $e_\mathrm{c} \approx 0$.
We thus adopt $e_\mathrm{c} = 0$ in the following stability analyses.

\begin{figure}
    \centering
	\includegraphics[width=\textwidth]{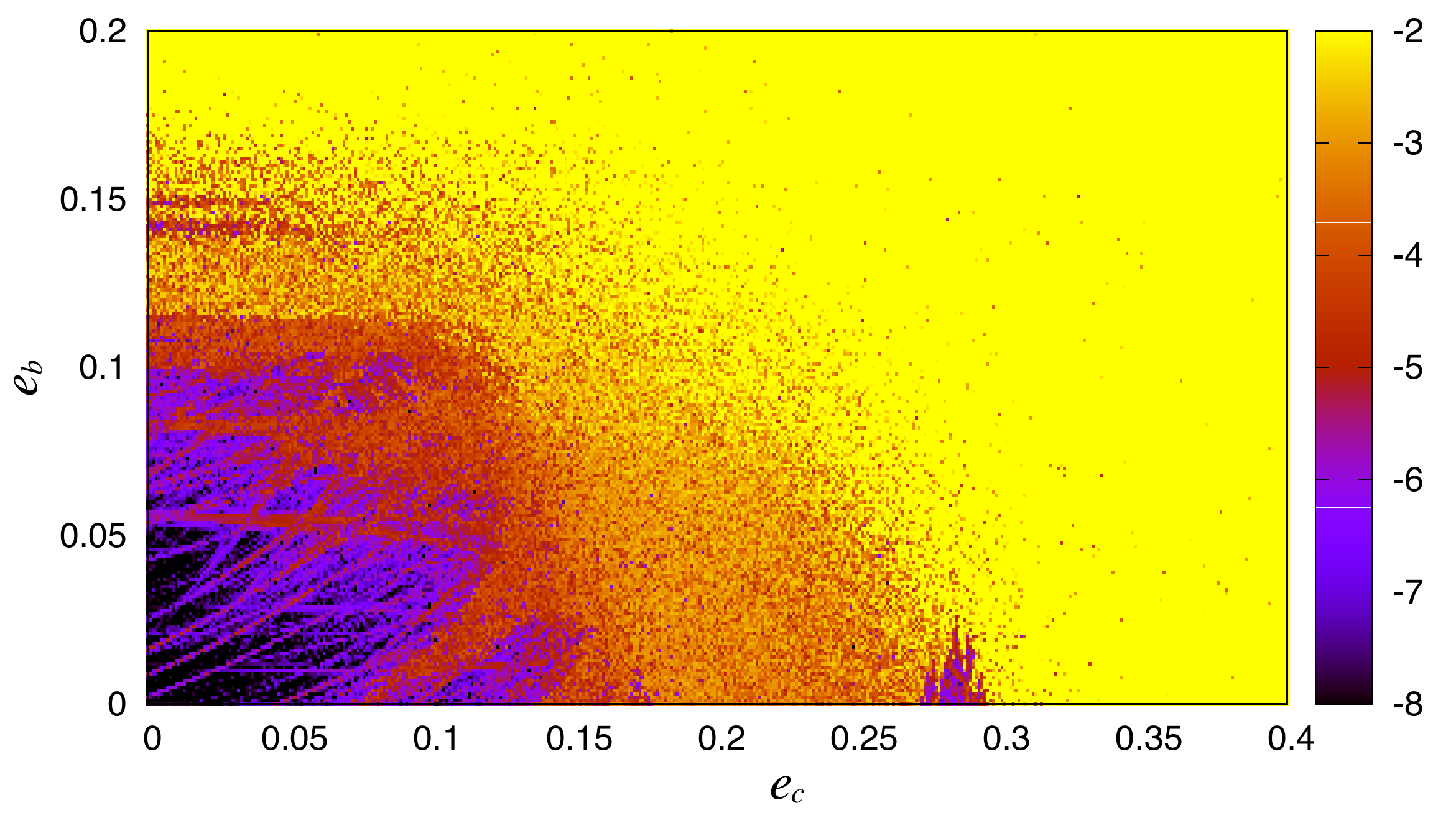}
    \caption{Stability analysis of the TOI-1338/BEBOP-1 system in the plane $(e_\mathrm{c}, e_\mathrm{b})$, assuming coplanar orbits. For fixed initial conditions (Table~1 main text), the parameter space of the system is explored by varying the eccentricities of both planets, with a step size of $0.001$. For each initial condition, the system is integrated over 5000~yr and a stability indicator is calculated which involved a frequency analysis of the mean longitude of the inner planet (see text). Chaotic diffusion is indicated when the main frequency of the mean longitude varies in time. Yellow points correspond to highly unstable orbits, while purple points correspond to orbits which are likely to be stable on a billion-years timescales.}   
    \label{fig:ecc}
\end{figure}

The inner planet is not detected in the radial-velocity data, its presence can only be inferred from photometric measurements \cite{Kostov2020}.
Therefore, in a second experiment, we explore the stability of the inner planet orbit, by varying the orbital period and the eccentricity of this planet, assuming coplanar orbits (Figure~12).
This allow us to test the compatibility of the two independent observational datasets.
We observe that the inner planet lies in a small stability region that is shaped by the presence of the binary system on the left hand side and by the presence of the massive outer planet on the right hand side.
%In the absence of the outer planet, all regions in red would become stable (see Fig.~14 in  \cite{Kostov2020}).
We thus conclude that the two planet circumbinary solution is reliable.
However, we assumed $e_\mathrm{c}=0$ to draw the stability map obtained in Figure~12.
As we increase the eccentricity of the outer planet, the small stable region is quickly degraded and completely disappears for $e_\mathrm{c} > 0.1$, in conformity with the results shown in Figure~11.
Figure~12 shows that the inner planet, TOI-1338/BEBOP-1b, is surrounded by unstable regions. If the planet migrated to its current location, it would have to pass through two resonances with the outer planet. This could indicate consequences for formation migration in the system.

\begin{figure}
    \centering
	\includegraphics[width=\textwidth]{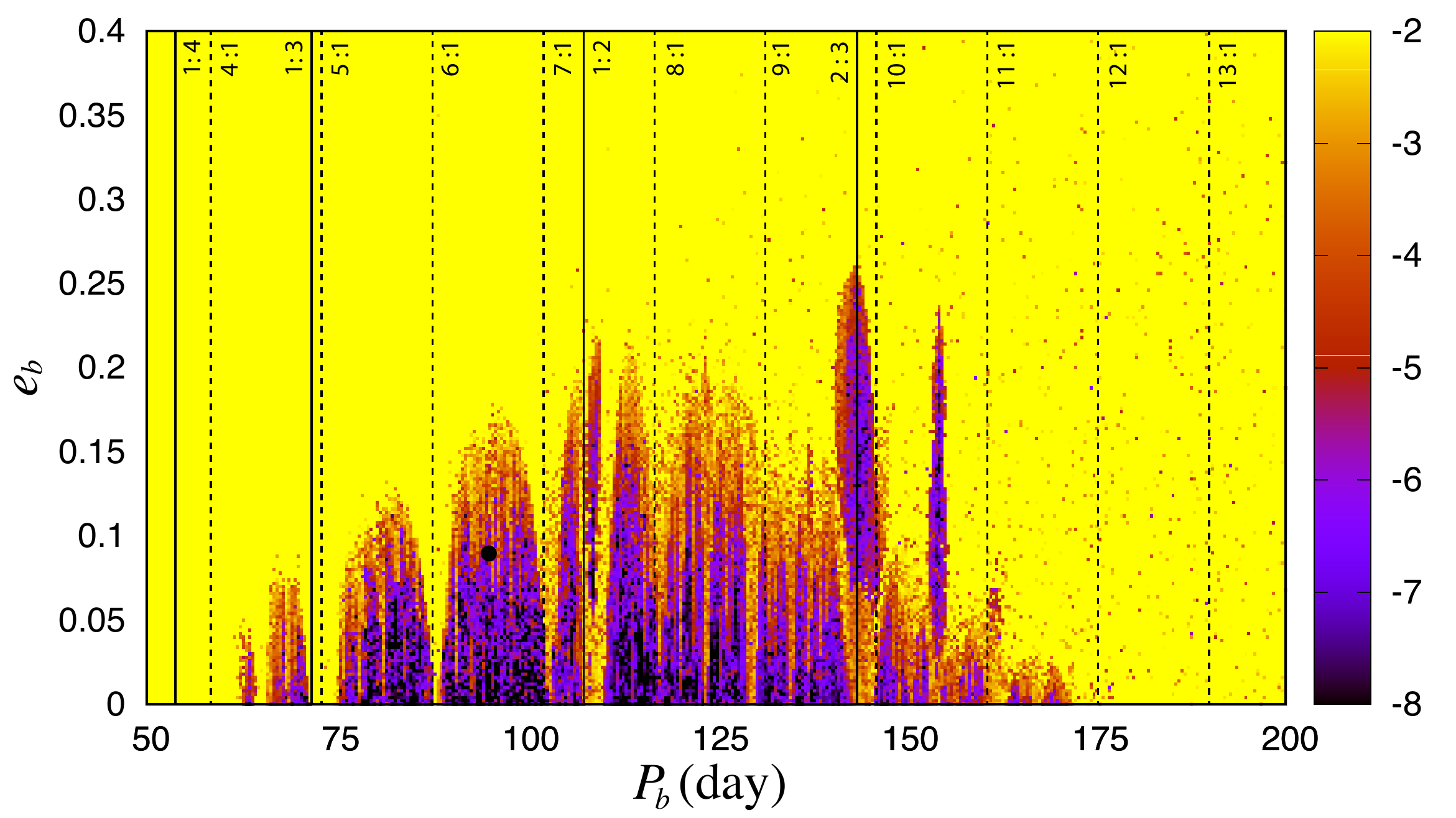}
    \caption{Stability analysis of the TOI-1338/BEBOP-1 system in the plane $(P_\mathrm{b}, e_\mathrm{b})$, assuming coplanar orbits. For fixed initial conditions (Table~1 main text), the parameter space of the system is explored by varying the orbital period $P_\mathrm{b}$ and the eccentricity $e_\mathrm{b}$ of the inner planet. The step size is $0.4$~day in orbital period and $0.002$ in eccentricity. The black dot gives the best fit parameters for the inner planet \cite{Kostov2020}. The color codes are the same from Figure~11, the dashed black lines indicate resonances with the binary, and the solid black lines resonances with the outer planet TOI-1338/BEBOP-1c.}   
    \label{fig:per}
\end{figure}

The radial-velocity technique alone is unable to constrain the inclination, $I_\mathrm{c}$, and the longitude of the node, $\Omega_\mathrm{c}$, of the outer planet.
As a result, we can only determine the minimum value of its mass, which corresponds to a coplanar system ($I_\mathrm{c}=90$), that we have been assuming in previous analyses.
However, contrarily to the inner planet, the outer planet does not transit, and thus cannot be exactly in the same orbital plane of the remaining bodies in the system.
In a final experiment, we explore the stability of the system, by varying the inclination and the longitude of the node of the outer planet (Figure~13).
We observe that the system can be stable within a circle centred at the coplanar solution, which corresponds to mutual inclinations smaller than $40^\circ$.
A larger stability regions also exists centred at ($\Omega_\mathrm{c} = 180^\circ, I_\mathrm{c}=90^\circ$), but it corresponds to mutual inclinations higher than $120^\circ$, that is, to retrograde orbits, which are more unlikely from a formation point-of-view.
As we change $I_\mathrm{c}$, the mass of the outer planet increases.
At the boundary of stability, $I_\mathrm{c} = 90^\circ \pm 40^\circ$, we get a maximum mass for the outer planet, $m_{\rm max, c} = 0.28$~M$_\mathrm{Jup}$.
Note, however, that we can also get a mutual inclination of $40^\circ$ with $I_c= 90^\circ$ and $\Omega_c = \pm 40^\circ$, for which the mass of the planet is on the low side, $m_{\rm min, c} = 0.21$~M$_\mathrm{Jup}$. Thus, the source of instability is truly mutual inclinations above the $40^\circ$ threshold and not masses higher than $0.28$~M$_\mathrm{Jup}$.
Again, we assumed $e_\mathrm{c}=0$ to draw the stability map obtained in Figure~12.
As we increase the eccentricity of the outer planet, the radius of the stable circle quickly shrinks and completely disappears for $e_\mathrm{c} > 0.1$, in conformity with the results shown in Figure~11.

\begin{figure}
    \centering
	\includegraphics[width=\textwidth]{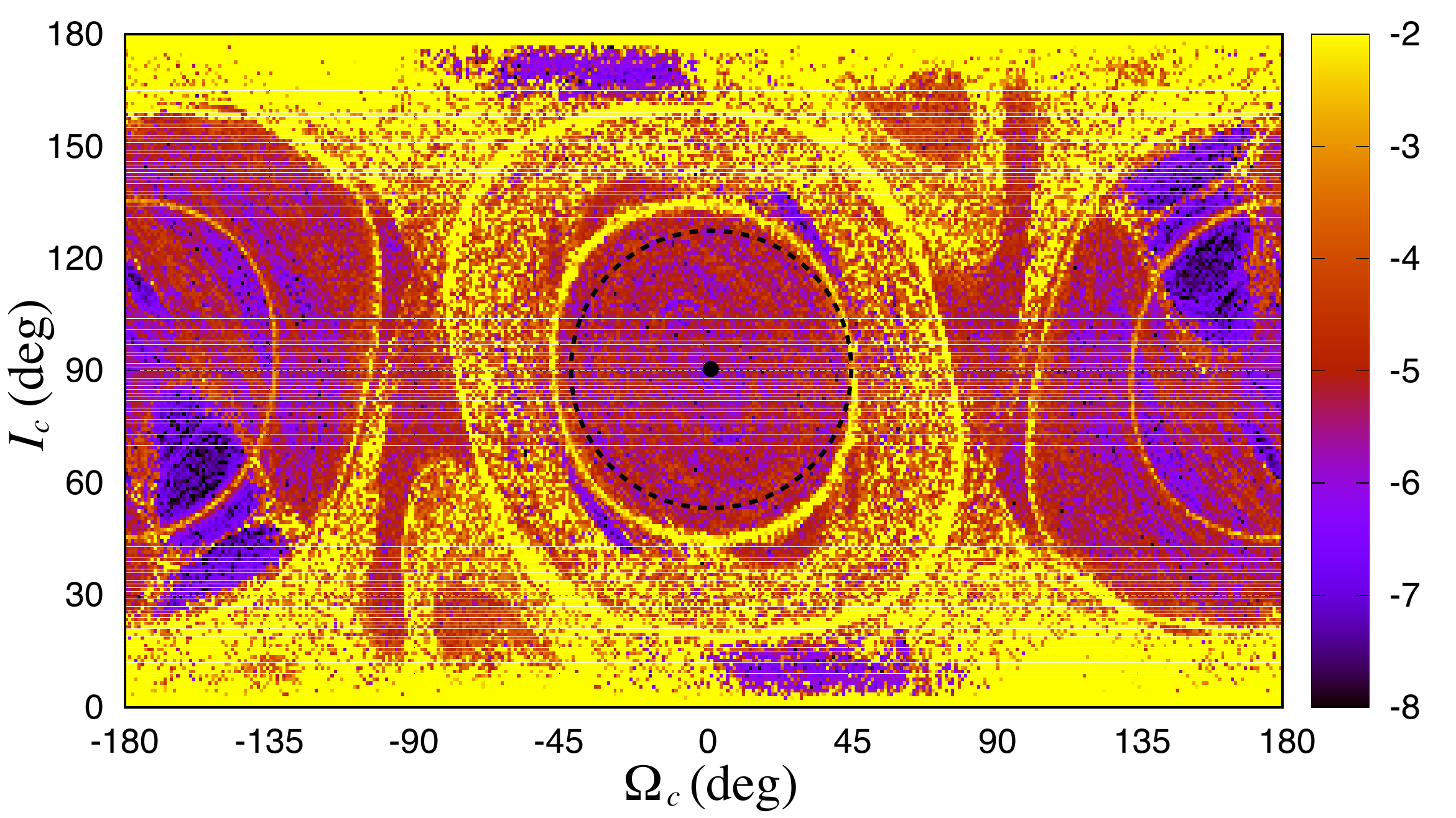}
    \caption{Stability analysis of the TOI-1338/BEBOP-1 system in the plane $(\Omega_\mathrm{c}, I_\mathrm{c})$. For fixed initial conditions (Table~1 main text), the parameter space of the system is explored by varying the inclination $I_\mathrm{c}$ and the longitude of the node $\Omega_\mathrm{c}$ of the outer planet, with a step size of $1^\circ$. The black dot gives the best fit parameters, while the dashed circle corresponds to mutual inclinations equal to $40^\circ$. The color codes are the same from Figure~11.}   
    \label{fig:inc}
\end{figure}

\subsection{Possible transits of TOI-1338/BEBOP-1c}\label{subsec:transitability}

Orbits of circumbinary planets exhibit nodal precession. This changes the orientation of the planet's orbit with respect to both the binary and the observer. This means that a planet changes from a transiting to non-transiting configuration \cite{Schneider1994,Martin&Triaud2014}. The precession timescale is typically on the order of decades, and hence can be observable with long baselines such as Kepler and TESS. Kepler-413 exhibited this on-off transit sequence \cite{Kostov2014}. Kepler-453 did not start transiting until half way through the Kepler mission \cite{Welsh2015}.

\cite{Martin&Triaud2015,Martin2017b} derived an analytic criterion for if a planet will ever enter transitability on the primary (A) or secondary (B) star during its precession cycle:

\begin{equation}
\label{eq:transitability}
    \Delta I > \left|\frac{\pi}{2} - I_{\rm bin} \right| - \sin^{-1}\left(\frac{a_{\rm A,B}}{a_{\rm p}} \sin \left|\frac{\pi}{2} - I_{\rm bin} \right| + \frac{R_{\rm A,B}}{a_{\rm p}} \right),
\end{equation}
where $\Delta I$ is the mutual inclination between the planet and binary orbit. It may seem counter-intuitive at first that a planet-binary misalignment makes transitability {\it more} likely. The reason is that the planet's sky inclination oscillates around the binary's sky inclination, with an amplitude equal to $\Delta I$. A higher $\Delta I$ therefore makes it more likely that the planet's inclination will become close to $90^{\circ}$, and hence transitability. For eclipsing binaries $I_{\rm bin}$ is near $90^{\circ}$, and hence the vast majority of circumbinary planets orbiting eclipsing binaries will eventually transit.

Placing the TOI-1338/BEBOP-1c values into Eq.~2 yields $\Delta I_{\rm min} = -0.11^{\circ}$ for the primary star. The negative inclination criterion means that TOI-1338/BEBOP-1c is {\it guaranteed to eventually transit} regardless of what its mutual inclination is. This is mainly because the binary is so well-aligned with our line of sight ($I_{\rm bin}=89.658^{\circ}$) and the primary star is rather large ($R_{\rm A}=1.299R_{\odot}$). Whilst we think that TOI-1338/BEBOP-1c is guaranteed to {\it eventually} transit, we are unable to predict when and how frequently. The precession period is

\begin{equation}
    \label{eq:precession_period}
    P_{\rm prec} = P_{\rm p}\frac{16}{3}\left(\frac{a_{\rm p}}{a_{\rm bin}} \right)^2\frac{1}{\cos\Delta I},
\end{equation}
where we assume circular orbits \cite{Schneider1994,Farago2010}. For TOI-1338/BEBOP-1c $P_{\rm prec} = 119$ years, assuming $\Delta I$ is close to zero, as is the case of the known circumbinary planets. With modern advances in medical science, there is a chance that the authors will live to see TOI-1338/BEBOP-1c transit.

% \subsection{The Habitable Zone}\label{subsec:hz}

% \textbf{David}: I'll put a section here about the habitable zone

\section{System formation}
The coplanarity between the planet and binary orbit planes for the circumbinary planets that have been discovered so far strongly suggests that these planets were formed in circumbinary protoplanetary discs that were themselves closely aligned with the binary orbit plane. We present here the results from a suite of simulations of circumbinary planet formation that were performed to examine plausible scenarios for the origin of the TOI-1338/BEBOP-1b \& c system.
\subsection{Circumbinary disc and N-body model}
Our simulations were performed using a newly developed code designed specifically to examine the formation of circumbinary planet systems. The code employs the N-body symplectic integrator \textsc{mercury6}, adapted to include a central binary system \cite{Chambers02,Chambers99}. The code utilises the `close-binary' algorithm \cite{Chambers02} that calculates the temporal evolution of the positions and velocities of each body in the simulations with respect to the centre of mass of the binary stars, subject to gravitational perturbations from both stars and other large bodies.
The evolution of the circumbinary disc is calculated using a 1D viscous $\alpha$--disc model \cite{Shakura73}, solving the standard diffusion equation with additional terms that account for the possible presence of a gap forming planet and a photoevaporative wind:
\begin{equation}
\label{eq:diffusion}
\dfrac{d\Sigma}{dt}=\frac{1}{r}\dfrac{d}{dr}\left[3r^{1/2}\dfrac{d}{dr}\left(\nu\Sigma r^{1/2}\right)-\dfrac{2\Lambda\Sigma r^{3/2}}{GM_{*}}\right]-\dfrac{d\Sigma_{\rm pe}}{dt}
\end{equation}
Here, $\dfrac{d\Sigma_{\rm pe}}{dt}$ is the rate of change in surface density due the photoevaporative wind, $\Lambda$ is the torque per unit mass that operates when a planet forms and becomes massive enough to open a gap in the disc, and $\nu$ is the disc viscosity \cite{Shak}, given by
\begin{equation}
\label{eq:viscosity}
\nu=\alpha c_{\rm s}^2/\Omega,
\end{equation}
where $c_{\rm s}$ is the local isothermal sound speed, $\Omega = \sqrt{\frac{GM_*}{r^3}}$ is the Keplerian frequency and $\alpha$ is the viscosity parameter.
The planet torque per unit mass is given by
\begin{equation}
    \Lambda = {\rm sign}(r-r_{\rm p})q^2\dfrac{GM_{*}}{2r}\left(\dfrac{r}{|\Delta_{\rm p}|}\right)^4,
\end{equation}
where $q$ is the planet/star mass ratio, $r_{\rm p}$ is the planet orbital radius, and $|\Delta_{\rm p}| = \max (H\text{, }|r-r_{\rm p}|)$, where $H$ is the local disc scale height.
We assume that the disc is in thermal equilibrium, and as such we use an iterative method to solve the following equation \cite{Dangelo12}
\begin{equation}
\label{eq:temperature}
Q_{\rm irr,A} + Q_{\rm irr,B} + Q_{\nu} + Q_{\rm cloud} - Q_{\rm cool} = 0,
\end{equation}
where we balance irradiation heating from both central stars ($Q_{\rm irr,A + B}$), background heating from the residual molecular cloud ($Q_{\rm cloud}$), viscous heating ($Q_{\nu}$) with blackbody cooling ($Q_{\rm cool}$).

Mass loss due to photoevaporative winds, resulting from high-energy photons emitted by both the central stars and nearby externally located stars, occurs on a time-scale determined by the adopted flux of high energy photons \cite{Dullemond07,Matsuyama03}.

Hydrodynamical simulations have shown that circumbinary discs develop precessing eccentric inner cavities, with their sizes and eccentricities being dependant on the properties of the binary and disc \cite{Pierens13,Thun17}. This cavity plays an important role in the evolution of planets that migrate into its vicinity. To mimic the presence of this cavity in the 1D viscous disc model, we use a variable $\alpha$ model fitted to results from 2D hydrodynamic simulations using \textsc{fargo3d} \cite{FARGO-3D-2016}, that show that the mass flux through the cavity remains roughly constant within the region around the binary. For these \textsc{fargo3d} simulations we adopt a computational domain that spans 0.13--20 $\au$, and for the numerical grid we adopt a resolution of $N_{\rm r} \times N_{\rm \phi} = 768 \times 512$, with logarithmic radial spacing. We use the stellar parameters outlined in this work, and the disc mass lying within 40 $\au$ was set to equal 6\% of the combined binary mass. The viscosity parameter $\alpha=10^{-3}$ and the constant disc aspect ratio was set to 0.05. To ensure the simulations reached equilibrium, we ran them for 30,000 binary orbits.
% The evolution of the circumbinary disc is calculated using a 1D viscous $\alpha$--disc model \cite{Shakura73}, where the equilibrium temperature is calculated by balancing irradiation heating from the central stars, background heating from the residual molecular cloud, viscous heating and blackbody cooling. Hydrodynamical simulations have shown that circumbinary discs develop precessing eccentric inner cavities, with their sizes and eccentricities being dependant on the properties of the binary and disc \cite{Pierens13,Thun17}.
% To mimic the formation of this cavity, we use a variable $\alpha$ model fitted to results from 2D hydrodynamic simulations using \textsc{fargo3d} \cite{FARGO-3D-2016}, that show that the mass flux through the cavity remains roughly constant within the region around the binary. Our approach to accounting for the precession and eccentricity of this cavity is described below.

\subsection{Planet migration}
Planets with masses that significantly exceed a Lunar-mass undergo migration because of gravitational interactions with the surrounding disc. We follow \cite{ColemanNelson14} and include Type I migration in the model via the culmination of the torque formulae that take account of Lindblad and corotation torques \cite{Paardekooper11}, as well as those of eccentricity and inclination damping \cite{Daisaka,cressnels},
\begin{equation}
\label{eq:typeItorque}
\begin{split}
&\Gamma_{\rm I,tot}=F_L\Gamma_{\rm LR}+\left\lbrace\Gamma_{\rm VHS}F_{p_v}G_{p_v}
+\Gamma_{\rm EHS}F_{p_v}F_{p_{\chi}}\sqrt{G_{p_v}G{p_{\chi}}}+\Gamma_{\rm LVCT}(1-K_{p_v})\right.\\
&\left.+\Gamma_{\rm LECT}\sqrt{(1-K_{p_v})(1-K_{p_{\chi}})}\right\rbrace F_e F_i
\end{split}
\end{equation}
where $\Gamma_{\rm LR}$, $\Gamma_{\rm VHS}$, $\Gamma_{\rm EHS}$, $\Gamma_{\rm LVCT}$ and $\Gamma_{\rm LECT}$, 
are the Lindblad torque, vorticity and entropy related horseshoe drag torques, and linear vorticity and 
entropy related corotation torques, respectively, as given by equations 3-7 in \cite{pdk11}. 
The functions $F_{p_v}$, $F_{p_{\chi}}$, $G_{p_v}$, $G_{p_{\chi}}$, $K_{p_v}$ and $K_{p_{\chi}}$ are 
related to the ratio between viscous/thermal diffusion time scales and horseshoe libration/horseshoe U-turn 
time scales, as given by equations 23, 30 and 31 in \cite{pdk11}.
If a planet becomes massive enough to form a gap in the disc \cite{Crida06}, its migration changes from type I to type II. When transitioning from type I to type II migration the model accounts for gap formation self-consistently by calculating the torque acting on the disc due to the planet, with the back-reaction driving migration \cite{Lin1986}. 
Planets in the vicinity of the inner cavity of the circumbinary disc will experience time varying forces arising from its eccentricity and precession. These are not naturally captured in a 1D disc model, so we have added additional terms to the evolution equations to incorporate these effects.
Using the results from the \textsc{fargo3d} simulations described above, we calculate an azimuthally averaged eccentricity profile, that when used in conjunction with a uniformly precessing disc, can give the velocity of the gas in an eccentric disc at any azimuthal location. Following the calculated precession rates from previous works \cite{Thun17}, we assume that the disc precession rate is equal to 3,000 binary orbits.
Using the steady precessing, 2D eccentric structure of the disc obtained from the hydrodynamic simulations, we created a two-dimensional map of the radial and azimuthal forces that an embedded particle at any position in the disc would feel. This map is interpolated during the N-body simulations to find the force acting on any planet that forms.
%To account for this, the model includes additional forces that are obtained by interpolating from a force map that has been generated from 2D hydrodynamic simulations of a circumbinary disc with binary parameters equal to those inferred for the BEBOP-1 system. These were performed using \textsc{fargo3d} \cite{FARGO-3D-2016}.

\subsection{Planetary growth}
The growth of planetary seeds in the model occurs via pebble accretion, and once a planet exceeds an Earth mass it starts to accrete gas from the disc.
Pebble accretion models have shown that as protoplanetary discs evolve, dust coagulates into pebbles and settles in the midplane, where the pebbles drift inwards \cite{Lambrechts14}.
This creates a pebble production front that expands outwards over time. As the pebbles drift through the disc, they encounter planetary embryos which can accrete the pebbles with an efficiency that depends on the disc and planet properties \cite{Lambrechts2012,Johansen17}.
We adopt the models of \cite{Lambrechts14,Johansen17} to account for the production of pebbles in the disc, as well as accretion onto planetary embryos.
Whilst such models were derived for single stars, with no perturbing forces from a central binary, they should be sufficiently accurate for the circumbinary disc models we present here. This is because the system is a close binary, and as such once material is only $\sim$few $\au$ away from the central stars the gravitational force it feels is very similar to that of a single star. Given that pebble production primarily occurs far from the binary, in the outer disc regions, we assume that the formulation for single stars is accurate.
In terms of planets accreting pebbles, the same applies with most planets accreting pebbles at distances greater than $2\au$.
However, some planets are able to migrate and accrete pebbles near the cavity, and here it would not be suitable to assume that the pebbles follow circular Keplerian orbits.
Results from previous work have shown that dust particles with small Stokes numbers closely follows the gas \cite{Coleman22b}, and as such we assume that the pebbles follow the same eccentric profiles as used to calculate the torques acting on the planet in the vicinity of the central cavity. The relative velocities between pebbles drifting past the planet are then calculated and used in the equations for pebble accretion \cite{Johansen17}.

Gas accretion onto the planets is modelled using a recently developed empirical fit \cite{Poon21} to the results of detailed 1D envelope structure models \cite{Coleman17} that account for local disc conditions:

\begin{align}\label{eq:gasenvelope-gc}
\left (\dfrac{d M_{\mathrm{ge}}}{d t} \right )_{\mathrm{local}}=& 10^{-10.199} \left( \dfrac{\mathrm{M_{\oplus}}}{\mathrm{yr}}\right) f_{\mathrm{opa}}^{-0.963} \left ( \dfrac{T_{\mathrm{local}}}{\mathrm{1\,K}}\right )^{-0.7049}\left(\dfrac{M_{\mathrm{core}}}{\mathrm{M}_{\oplus}} \right )^{5.6549}  \left (\dfrac{M_{\mathrm{ge}}}{\mathrm{M}_{\oplus}}  \right )^{-1.159}\nonumber \\
&\times\left [ \exp{\left ( \dfrac{M_{\mathrm{ge}}}{M_{\mathrm{core}}} \right )} \right ]^{3.6334}.
\end{align}
where $T_{\rm local}$ is the local disc temperature, $f_{\rm opa}$ is an opacity reduction factor, equal in this work to 0.01, and $M_{\rm core}$ and $M_{\rm ge}$ are the planet's core and envelope masses respectively.
Given that these models calculate orbit averaged accretion rates, we assume this is also valid for planets in eccentric circumbinary discs, however further work examining the validity of this assumption should be undertaken.
We note these models predict that gas accretion starts off very slowly and speeds up as the core and envelope grow in mass. Hence, the onset of gas accretion does not generally result in the formation of a gas giant planet because the envelope contracts on its Kelvin-Helmholtz time-scale.

\subsection{The resulting circumbinary planet systems}
The simulations were initiated by placing 42 planetary seeds throughout the disc between 2 and 20~au, with eccentricities and inclinations distributed uniformly up to $e=0.02$ and $i=0.5^{\circ}$, respectively.
The initial masses of the seeds were set to equal one tenth of the transition mass, which is where pebble accretion becomes efficient as the planetary core begins to accrete from the entirety of its Hill sphere as opposed to its Bondi sphere. These initial masses were between $10^{-4}$ and $10^{-3} M_{\oplus}$ and are consistent with recent simulations of planetary embryo formation within protoplanetary discs based on gravitational collapse \cite{Coleman21}.
We initialised the gas surface density according to $\Sigma=\Sigma_0\left(\frac{r}{1\au}\right)^{-1}$ where $\Sigma_0$ depends on the initial disc mass.
We explored discs masses between 10--15\% of the combined stellar mass in discs that extend out to 100$\au$, and we considered metallicities of 0.5, 1 or 2 $\times$ the Solar value. We assumed a viscous $\alpha$ of $2\times10^{-3}$, an $f_{41}$ parameter for EUV photoevaporation of 100, and an external photoevaporative mass loss rate for a disc of size 100$\au$ of $3\times10^{-7}\msunyr$.

Figure~14 shows the mass and semimajor axis evolution from an example simulation that formed a planetary system similar to TOI-1338/BEBOP-1.
Black points show the final masses and semimajor axes of the surviving planets, whilst the vertical black dashed line denotes the stability boundary \cite{Holman1999}. The red triangle and plus sign indicate the locations of TOI-1338/BEBOP-1b and c.

The formation of this system occurred as follows.
As the pebble front moved outwards, the planetary embryos on the most circular orbits were able to accrete pebbles efficiently, allowing a number of them to grow to masses greater than an Earth mass.
These planets began to migrate, generally in towards the central cavity, and continued to accrete drifting pebbles as well as gas. The central cavity in the disc provides a positive surface density gradient in the gas, and this allowed corotation torques to balance Lindblad torques for planets arriving there, resulting in migration stalling.
Hence, inward migrating planets started to congregate near the outer edge of the cavity, with some resonant chains forming.
As the planets continued to grow in mass, two planets with masses $\sim 10 M_{\oplus}$ collided, forming a more massive core.
This planet was then able to accrete gas more efficiently and entered a brief period of runaway gas accretion to become a gas giant.
As the planet's mass increased, it migrated in closer towards the central stars, pushing planets interior to it into the cavity, and closer to the stability boundary.
This resulted in the inner-most planets being ejected from the system as they crossed the stability boundary and were gravitationally scattered by the central stars, although the planet destined to become the TOI-1338/BEBOP-1b analogue remained outside of the boundary. The gas giant planet opened a gap in the disc and began to undergo slow inwards type II migration, whilst continuing to accrete gas through the gap at the viscous supply rate. After 1.8~Myr the disc reached the end of its life and was fully dispersed due to photoevaporative winds, and the giant planet stopped accreting and migrating, leaving both planets with masses and semimajor axes similar to those inferred from the observations.
The system was allowed to evolve for a further 8 Myr, during which time it remained stable, yielding the final system shown in figure~14.

This evolution scenario was a common outcome for a number of simulated systems, with planets growing through pebble accretion at large radii in the disc, before migrating in towards the central cavity. Planets congregated there, leading to collisions and some ejections.
Figure~15 shows the mass versus semimajor axis for all planets formed in the simulations, with the colour coding showing the metallicity of the system in comparison with the Solar value (this determines the mass in pebbles assumed in the models).
Planets shown by grey points have been lost from the systems through collisions with more massive planets or by ejection.
The black dashed line shows the stability boundary, whilst the black triangle and plus sign show the masses and locations of TOI-1338/BEBOP-1b and c.

The figure shows that formation of TOI-1338/BEBOP-1b analogues is a relatively common outcome of the simulations, with numerous super-Earth and Neptune mass planets being formed with similar orbital periods.
This is unsurprising given that these planets become trapped relatively easily around the cavity, close to the observed location of TOI-1338/BEBOP-1b.
For TOI-1338/BEBOP-1c, this planet has fewer simulated planets in its proximity, a result of its observed mass.
At this mass, planets are in a runaway gas accretion regime, where they quickly grow to masses greater than Saturn's.
Indeed, in our best fit simulation, the gas giant planet had a mass roughly twice that inferred for TOI-1338/BEBOP-1c.
In our best fitting simulations, runaway gas accretion was initiated close to the end of the disc lifetimes, which limited the amount of gas that could be accreted by the planets.
This necessity for appropriate timing could indicate that TOI-1338/BEBOP-1c formed very late in the disc lifetime.

Looking at the metallicities required to form the planets, no systems with half the Solar metallicity were able to form systems similar to TOI-1338/BEBOP-1.
This was because cores with sufficient mass to undergo efficient gas accretion were unable to form. On the other hand, systems with twice the Solar metallicity were too effective at forming massive cores, resulting in too many gas giant planets forming, and few planets with masses comparable to TOI-1338/BEBOP-1b remaining in the system, making them poor fits for comparison.
These results show that the metallicity of the system, and how efficiently dust is converted into planets, is important for determining the architectures of circumbinary planet systems.

Finally, we note that the orbital configuration of TOI-1338/BEBOP-1, with planet periods of 95 and 215 days and the more massive planet on the outside, is consistent with the prediction of Fitzmaurice et al. (2022) \cite{Fitzmaurice2022} in their study of migrating multi-planet circumbinary systems. They predict that if there is a planet near the stability limit, as is the case for TOI-1338/BEBOP-1, then any equal or more massive outer planets would be located on a period more than double that of the inner planet. Their simulations show that if an exterior planet carrying greater angular momentum reaches the 2:1 resonance with the inner planet, the two lock into resonance and the inner planet is pushed into the inner cavity, resulting in ejection. The periods in TOI-1338/BEBOP-1 at 95 and 215 days are explained by the disc dissipating (and hence stopping migration) before the outer planet reaches the 2:1 resonance.

\begin{figure}
    \centering
	\includegraphics[width=\textwidth]{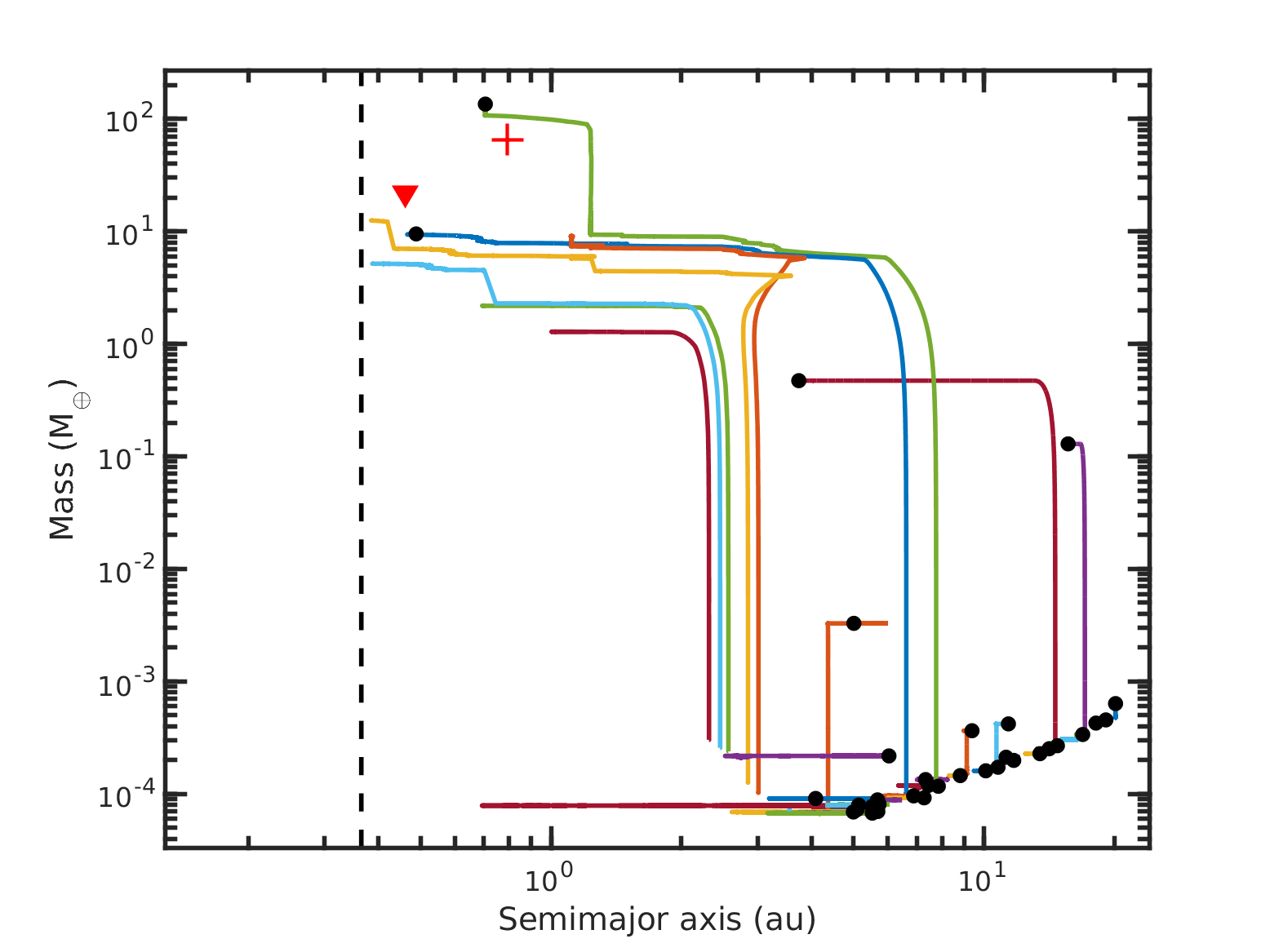}
    \caption{Evolution of the mass versus semimajor axes for the best matching simulated planetary system formed through pebble accretion to TOI-1338/BEBOP-1. Black dots represent the final masses and locations of the planets, whilst the red triangle and plus show the observed locations of TOI-1338/BEBOP-1b and c respectively. The black dashed line denotes the stability limit for the system \cite{Holman1999}.}   
    \label{fig:form}
\end{figure}

\begin{figure}
    \centering
	\includegraphics[width=\textwidth]{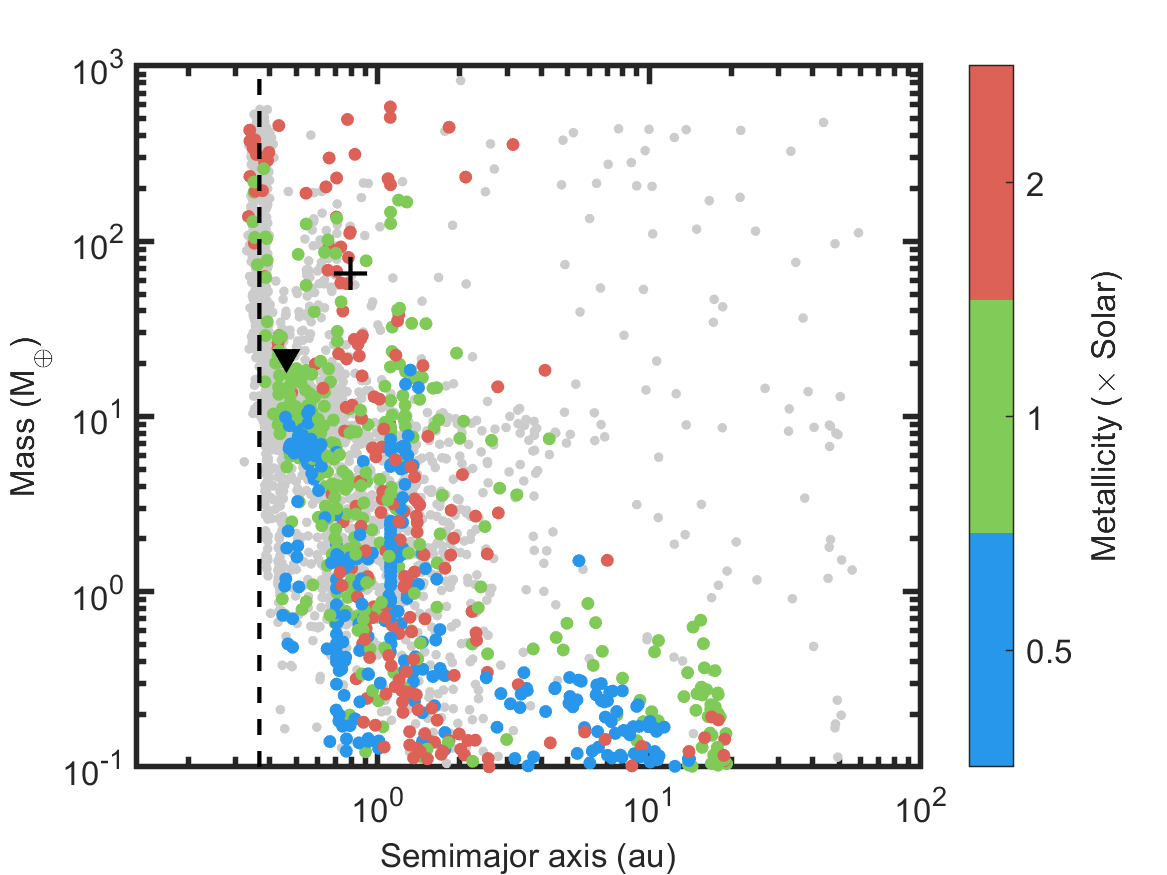}
    \caption{A mass versus semimajor axes plot showing all planets formed in simulations. Different colour points show the initial metallicity of the sytems, whilst the grey points show planets there were lost, either through accretion into other planets, or ejection from the system. TOI-1338/BEBOP-1b and 1c are denoted by the black triangle and plus symbols. The black dashed line denotes the stability limit for the system \cite{Holman1999}.}   
    \label{fig:all_mva}
\end{figure}

\section{Prospects for atmospheric follow-up with JWST}

In order to assess the suitability of TOI-1338/BEBOP-1b for atmospheric characterisation using transmission spectroscopy, we calculate its Transmission Spectroscopy Metric (TSM) as established by Kempton et. al. (2018) \cite{Kempton2018}. A planet's TSM is proportional to the SNR that could be achieved by observations using {\it JWST}, and it depends on the planet's mass, radius and equilibrium temperature, and the host star's radius and J-band magnitude.

Calculation of TOI-1338/BEBOP-1b's equilibrium temperature is non-trivial; its non-zero eccentricity and changing irradiation levels caused by the binary mean that the temperature oscillates between minimum and maximum values. Full modelling of this, as was done in Kane \& Hinkel (2013) \cite{Kane2013}, is beyond the scope of this paper, so we estimate a mean equilibrium temperature based on two extremes: when the secondary star is closest and furthest to the planet. The equilibrium temperature calculated from the irradiation of just the primary star is $\rm 501\,K$, and the motion of the secondary star will cause the true equilibrium temperature to vary between $\rm 603-659\,K$.

We calculate the TSM following Kempton et al. (2018) \cite{Kempton2018} using a mean $\rm T_{eq}=630\,K$ and the mass upper limit of $\rm M_b\,<\,21.8\,M_{\oplus}$ in the first instance, yielding a TSM of 39. This value is below the suggested JWST cut-off of 96 for planets in the sub-Jovian regime. However, the mass used for this calculation is the upper limit; a lower mass would imply a larger scale height and therefore a significantly larger TSM. 

In Figure~16 we present the TSM of TOI-1338/BEBOP-1b as compared to other known exoplanets. Its position shows that even the lower limit of its TSM range compares favourably with other sub-Jovian sized planets in its temperature regime. We have indicated with an arrow the range of TSMs possible for a mass down to $\rm 10\, M_{\oplus}$, which would imply a mean density of $\rm 0.17\, g/cm^3$, comparable to Kepler-47c, for example \cite{Orosz2019}.

We also note that crucially, of the now 15 known circumbinary exoplanets, TOI-1338/BEBOP-1b is the only one for which observations of this kind can currently be pursued. Most Kepler systems are too faint (e.g. Kepler-47 with a $\rm Jmag=13.970$ \cite{2MASSCat}, or Kepler-34 with a $\rm Jmag=13.605$ \cite{2MASSCat}), and others such as Kepler-16 no longer transit \cite{Doyle2011}. Therefore, despite the challenges it may present, TOI-1338/BEBOP-1b is our only possibility to shed light on the atmospheric make-up of circumbinary planets.

\clearpage
\begin{figure}[ht!]
    \centering
    \includegraphics[width=\textwidth]{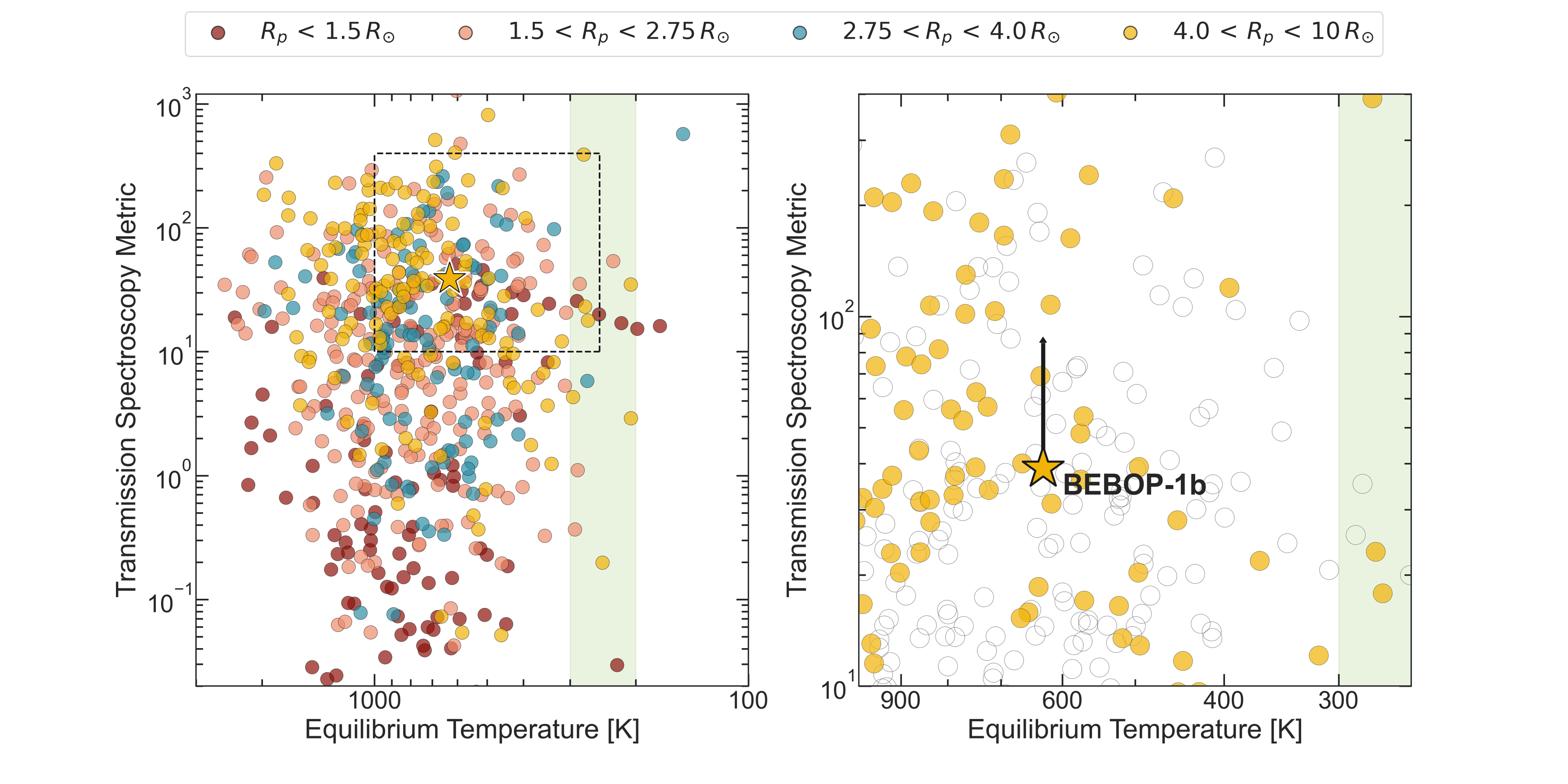}
    \caption{Transmission spectroscopy metric as a function of equilibrium temperature for planets with measured masses. Planets are coloured according to their size range, with TOI-1338/BEBOP-1b plotted as a yellow star as it falls into the sub-Jovian regime of $\rm 4.0<R_p<10~R_{\oplus}$. In both panels we show in light green the position of the temperate zone between $\rm 200-300~K$. In the right hand panel we zoom in on the vicinity of TOI-1338/BEBOP-1b and indicate with a black arrow the range of possible TSMs for a mass down to $\rm 10~M_{\oplus}$. Data for this plot was retrieved from the NASA Exoplanet Archive \cite{ExoArchive} on May 18th 2022.}
    \label{fig:TSM}
\end{figure}

%%%%%%%%%%%%%%%%%%%% REFERENCES %%%%%%%%%%%%%%%%%%

% The best way to enter references is to use BibTeX:

\bibliography{BEBOP-1}

\bibliographystyle{Science}

%%%%%%%%%%%%%%%%%%%% RV data %%%%%%%%%%%%%%%%%%

\section{Journal Of Radial-Velocity Observations}

\onecolumn
\begin{longtable}[1]{cllllll}\label{tab:BEBOP-1_HARPS_data}\\
	\caption{Journal of Observations containing our HARPS data for TOI-1338/BEBOP-1. Flags indicate whether the measurement is identified as a potential outlier for the following reasons: W, wrong star; B, bisector outlier; F, FWHM outlier; R, during binary transit; P, during planetary transit; S, identified by student-t. Dates are given in BJD - 2,400,000. $V_{\rm rad}$ are the measured radial velocities with their uncertainties $\sigma_{V_{\rm rad}}$. FWHM is the Full With at Half Maximum of the Gaussian fitted to the cross correlation function, and {\it contrast} is its amplitude. Bis. span is the span of the bisector slope. Uncertainties on FWHM and bis. span are $2\times\sigma_{V_{\rm rad}}$.}\\
	\hline
	\hline
    flag&BJD-2,400,000 &$V_{\rm rad}$&$\sigma_{V_{\rm rad}}$&FWHM&contrast&bis. span\\
    &[days]&[$\rm km\,s^{-1}$]&[$\rm km\,s^{-1}$]&[$\rm km\,s^{-1}$]& & [$\rm km\,s^{-1}$]\\
    \hline
    \endfirsthead
    \hline
	\hline
        flag&BJD&$V_{\rm rad}$&$\sigma_{V_{\rm rad}}$&FWHM&contrast&bis. span\\
	\hline
    \endhead
    \hline
    \multicolumn{5}{l}{Table continues on next page...}\\
    \hline
    \endfoot
    \hline
    \endlastfoot 
&58216.529441&50.78673&0.00527&8.62517&33.284&0.04984\\
&58436.821943&48.95425&0.00683&8.60645&33.510&0.01450\\
&58488.758689&23.57766&0.00433&8.61047&33.532&0.02913\\
&58503.731900&26.22766&0.00648&8.62748&32.803&0.04815\\
&58516.712301&14.71036&0.00592&8.62249&33.393&0.03321\\
&58556.646009&20.57355&0.00664&8.61148&33.356&0.02785\\
&58566.604454&50.14641&0.00526&8.65706&33.177&0.06616\\
&58698.935537&50.75479&0.00501&8.63891&33.166&0.04375\\
&58708.900586&30.92087&0.00844&8.62618&33.307&0.03230\\
&58717.869356&14.86234&0.00608&8.64292&33.309&0.03386\\
&58721.884268&19.14344&0.00577&8.63451&33.503&0.03270\\
&58726.880800&49.14747&0.00606&8.63034&33.580&0.02617\\
&58740.815851&46.73639&0.00673&8.65363&33.015&0.02828\\
&58777.865197&7.57447&0.00775&8.62193&33.480&0.04777\\
&58832.810367&37.48155&0.00573&8.63670&33.207&0.03620\\
&58848.722188&21.71885&0.00350&8.64730&33.414&0.02062\\
F&58861.667840&41.08977&0.00616&8.55803&33.640&0.03357\\
&58867.644660&16.88223&0.00783&8.64317&33.465&0.02665\\
&58902.582483&50.11217&0.00377&8.60631&33.474&0.04012\\
&59163.705508&43.13564&0.00417&8.60739&33.669&0.04133\\
&59170.792366&14.32580&0.00741&8.59433&33.516&0.02690\\
&59173.721395&12.42927&0.00439&8.61091&33.624&0.02420\\
&59176.668061&32.95458&0.00417&8.61199&33.595&0.01304\\
&59188.634700&14.25913&0.00636&8.62851&33.702&0.04388\\
&59191.651879&35.49221&0.00486&8.59691&33.561&0.02025\\
&59211.662441&45.90963&0.00487&8.61660&33.531&0.04283\\
&59227.640383&32.91789&0.00448&8.61194&33.672&0.03462\\
&59244.662941&8.79879&0.00444&8.64343&33.221&0.01978\\
&59251.644571&44.61290&0.00520&8.63634&33.503&0.02609\\
&59254.602429&49.90968&0.00448&8.63231&33.600&0.03055\\
&59257.658434&22.97524&0.00496&8.62586&33.461&0.02828\\
&59260.590560&8.80567&0.00614&8.63341&33.413&0.06138\\
&59272.597942&19.04391&0.00456&8.61572&33.388&0.02352\\
&59276.545689&15.95709&0.00645&8.65730&33.152&0.05425\\
&59280.570701&43.13379&0.00535&8.65045&33.420&0.05059\\
&59282.587090&50.43272&0.00429&8.66531&33.226&0.03972\\
&59286.640886&25.88951&0.00658&8.61449&33.288&0.04448\\
&59291.599947&19.03116&0.00717&8.67435&32.989&0.01348\\
&59296.574262&49.02190&0.00485&8.65725&33.296&0.05447\\
&59432.915546&23.53480&0.01006&8.60847&33.470&-0.00621\\
&59461.840683&27.18387&0.00837&8.63513&33.182&0.02817\\
F&59474.875056&44.09898&0.00608&8.55833&33.404&0.05304\\
&59530.787352&50.18431&0.00435&8.59874&33.507&0.04165\\
&59533.768571&40.07132&0.00398&8.59069&33.524&0.03166\\
W/S&59536.865577&8.58978&0.00536&8.78151&32.610&-0.05729\\
&59592.644734&35.36664&0.00442&8.58328&33.543&0.02524\\
&59610.628629&7.60999&0.00424&8.60636&33.508&0.03719\\
&59617.633579&47.93738&0.00585&8.62591&33.392&0.04730\\
&59638.669783&11.07926&0.00542&8.61874&33.062&0.02959\\
&59640.662829&9.73189&0.00551&8.62356&33.143&0.02188\\
&59643.549232&28.77716&0.00566&8.60709&33.489&0.04687\\
&59645.617218&42.22367&0.00624&8.62776&32.944&0.05350\\
&59648.619602&50.65155&0.00662&8.64878&32.875&0.01530\\
&59652.605979&17.00512&0.00764&8.58887&32.870&0.03801\\
&59658.559334&31.63510&0.00474&8.60176&33.121&0.01684\\
&59662.578359&50.64357&0.00709&8.61825&32.884&0.04475\\
&59666.562593&24.65893&0.00460&8.62847&33.322&0.02856\\
&59678.558881&48.91383&0.00613&8.63485&33.275&0.04387\\
&59682.528521&10.85815&0.00699&8.61124&33.206&0.07145\\
&59686.540476&22.71545&0.00689&8.60983&32.928&0.05059\\
&59688.495789&36.51633&0.00633&8.61060&33.306&0.04156\\
	\hline
\end{longtable}

\onecolumn
\begin{longtable}[1]{cllllll}\label{tab:BEBOP-1_ESPRESSO19_data}\\
	\caption{Journal of Observations containing our 2019 ESPRESSO data for TOI-1338/BEBOP-1 pre-calibration lamp change. Flags indicate whether the measurement is identified as a potential outlier for the following reasons: W, wrong star; B, bisector outlier; F, FWHM outlier; R, during binary transit; P, during planetary transit; S, identified by student-t. Columns are as in Table~3.}\\
	\hline
	\hline
    flag&BJD-2,400,000 &$V_{\rm rad}$&$\sigma_{V_{\rm rad}}$&FWHM&contrast&bis. span\\
    &[days]&[$\rm km\,s^{-1}$]&[$\rm km\,s^{-1}$]&[$\rm km\,s^{-1}$]& & [$\rm km\,s^{-1}$]\\
    \hline
    \endfirsthead
    \hline
	\hline
        flag&BJD&$V_{\rm rad}$&$\sigma_{V_{\rm rad}}$&FWHM&contrast&bis. span\\
	\hline
    \endhead
    \hline
    \multicolumn{5}{l}{Table continues on next page...}\\
    \hline
    \endfoot
    \hline
    \endlastfoot 
&58736.877282&21.76153&0.00386&9.35270&45.08543&-0.02995\\
&58740.853942&46.75387&0.00508&9.37866&43.97280&-0.02901\\
&58743.882022&47.54869&0.00525&9.34604&44.67442&-0.02851\\
F&58760.868104&24.05023&0.00525&9.40855&44.55251&-0.00934\\
&58774.797532&32.50228&0.00288&9.36375&45.12590&-0.00107\\
&58788.804499&39.20782&0.00281&9.35552&46.13411&-0.01414\\
&58800.687988&50.49684&0.00211&9.33485&46.21587&0.00094\\
&58805.840799&11.96096&0.00293&9.34411&46.18027&-0.02645\\
&58805.853819&11.86026&0.00276&9.32303&46.18795&-0.02347\\
&58812.832857&41.52667&0.00202&9.34686&46.40748&-0.00833\\
&58820.719656&10.11904&0.00191&9.33895&46.26380&-0.01669\\
&58831.662527&46.86291&0.00312&9.34385&45.91930&-0.00895\\
&58835.596434&8.75913&0.00302&9.35211&45.82418&-0.03332\\
&58839.611578&25.24170&0.00281&9.34291&45.97437&0.00664\\
&58843.641422&48.67236&0.00338&9.36552&45.89060&-0.03628\\
&58864.576277&9.94548&0.00245&9.33974&46.43191&-0.03564\\
&58880.582185&8.10059&0.00260&9.36042&46.02511&-0.03986\\
&58887.547492&48.91687&0.00226&9.34008&46.16192&-0.03410\\
&58895.570898&9.36537&0.00245&9.35557&45.63188&-0.02719\\
&58897.568595&21.74950&0.00285&9.33254&45.99687&-0.00162\\
	\hline
\end{longtable}

\onecolumn
\begin{longtable}[1]{cllllll}\label{tab:BEBOP-1_ESPRESSO21_data}\\
	\caption{Journal of Observations containing our 2021/22 ESPRESSO data for TOI-1338/BEBOP-1 post-calibration lamp change. Flags indicate whether the measurement is identified as a potential outlier for the following reasons: W, wrong star; B, bisector outlier; F, FWHM outlier; R, during binary transit; P, during planetary transit; S, identified by student-t. Columns are as in Table~3.}\\
	\hline
	\hline
    flag&BJD-2,400,000 &$V_{\rm rad}$&$\sigma_{V_{\rm rad}}$&FWHM&contrast&bis. span\\
    &[days]&[$\rm km\,s^{-1}$]&[$\rm km\,s^{-1}$]&[$\rm km\,s^{-1}$]& & [$\rm km\,s^{-1}$]\\
    \hline
    \endfirsthead
    \hline
	\hline
        flag&BJD&$V_{\rm rad}$&$\sigma_{V_{\rm rad}}$&FWHM&contrast&bis. span\\
	\hline
    \endhead
    \hline
    \multicolumn{5}{l}{Table continues on next page...}\\
    \hline
    \endfoot
    \hline
    \endlastfoot 
&59207.687480&43.80224&0.00156&9.37761&46.31836&-0.01166\\
S&59209.628417&50.37108&0.00159&9.37580&46.43171&-0.02890\\
&59211.660755&45.77950&0.00217&9.35553&46.17093&-0.03101\\
&59213.608767&25.62042&0.00147&9.35216&45.85679&-0.01713\\
&59216.617681&8.18971&0.00153&9.37519&46.34949&-0.03437\\
&59218.740865&20.18316&0.00159&9.36935&46.11973&-0.01806\\
&59226.586386&43.49514&0.00155&9.36541&46.26637&-0.03917\\
&59230.661220&7.43254&0.00144&9.36048&46.40324&-0.04251\\
F&59234.763082&30.43838&0.00253&9.38958&45.42103&-0.02132\\
&59237.675552&47.27860&0.00239&9.38360&45.80074&-0.01717\\
R&59242.594322&28.37619&0.00158&9.35821&46.07464&0.02823\\
&59251.647333&44.47071&0.00156&9.36148&46.26004&-0.00822\\
&59253.579655&50.52710&0.00157&9.36675&46.24797&-0.02805\\
&59255.616689&44.90922&0.00231&9.36382&45.55454&-0.01475\\
&59260.656954&8.86909&0.00349&9.38242&45.03587&-0.02082\\
&59261.570641&13.44921&0.00147&9.36278&46.13836&-0.01415\\
&59263.543084&27.30447&0.00174&9.35886&46.22777&-0.02054\\
&59265.633180&41.15591&0.00192&9.37520&45.80166&-0.03367\\
&59269.686062&48.06284&0.00402&9.36591&45.33246&-0.02726\\
&59270.617600&41.76652&0.00167&9.36544&46.13202&-0.00691\\
&59279.615948&37.38595&0.00148&9.32690&46.62544&-0.02920\\
&59284.600155&46.44112&0.00154&9.37661&46.35578&-0.00420\\
&59288.601602&8.24637&0.00393&9.39004&45.29095&-0.01497\\
&59289.558302&7.94851&0.00140&9.36444&46.36742&-0.01240\\
&59292.617105&26.25967&0.00160&9.36279&46.38443&-0.01914\\
&59298.529197&49.47896&0.00180&9.36071&46.01858&-0.01272\\
&59310.557405&46.63806&0.00277&9.35520&45.72558&0.01329\\
R&59315.529230&30.05916&0.00161&9.22765&46.59440&-0.01299\\
&59318.493789&7.49658&0.00334&9.37612&46.25952&0.00924\\
&59434.886191&7.72835&0.00406&9.38789&45.20792&-0.02862\\
&59461.875388&26.60478&0.00179&9.38405&46.00120&-0.00395\\
F&59464.797631&7.78951&0.00610&9.38198&44.24505&-0.03947\\
&59464.852082&7.91591&0.00562&9.43539&44.26999&-0.00213\\
&59466.891173&18.78338&0.00248&9.33472&45.93830&-0.02929\\
&59471.888348&48.91225&0.00257&9.35499&46.10455&-0.00382\\
&59473.770470&49.65291&0.00441&9.37446&45.85213&-0.02673\\
F&59475.781757&35.16473&0.00544&9.40688&44.61897&-0.03005\\
F&59475.793768&35.03023&0.00471&9.40944&45.39922&-0.02857\\
&59491.728251&18.88142&0.00603&9.31472&46.12284&-0.07435\\
&59493.720410&7.44845&0.00559&9.35603&45.28000&-0.04729\\
&59509.854678&13.07129&0.00236&9.34211&46.23157&-0.02061\\
&59512.842954&34.02253&0.00245&9.35205&46.87258&-0.00193\\
&59512.855927&34.11283&0.00263&9.35547&46.69308&-0.00158\\
&59529.677831&46.60567&0.00361&9.32951&45.79053&-0.02134\\
F&59530.618883&49.68125&0.00494&9.40343&44.94001&-0.02776\\
&59531.802019&50.44588&0.00226&9.33524&46.23785&-0.03232\\
&59532.720289&47.66931&0.00204&9.33130&46.39628&-0.01964\\
&59533.679007&40.78062&0.00228&9.34486&46.03277&-0.00715\\
R&59534.615899&30.40381&0.00337&9.35282&45.33094&0.01244\\
&59535.766807&16.55441&0.00255&9.34206&45.89073&-0.03407\\
F&59536.619244&9.64249&0.00294&9.41335&45.01306&-0.03384\\
F/S&59537.738391&7.59953&0.00371&9.41515&46.07402&-0.03767\\
&59542.613295&37.64360&0.00311&9.37607&45.61895&-0.04708\\
&59543.758054&44.17420&0.00190&9.33951&46.37724&-0.02434\\
&59544.665713&48.05618&0.00205&9.34654&46.12688&-0.01813\\
&59545.669437&50.44361&0.00231&9.35663&46.08663&-0.04244\\
&59546.842367&49.57590&0.00287&9.32741&46.09767&-0.03066\\
&59547.702845&45.48948&0.00291&9.36807&45.72661&-0.01271\\
&59548.622872&37.39873&0.00351&9.38229&45.23370&0.00187\\
&59549.670272&24.81471&0.00212&9.34777&45.99796&-0.00368\\
&59550.722645&13.21611&0.00169&9.32627&46.16202&-0.00265\\
&59551.671532&7.91486&0.00293&9.35368&45.91887&0.03150\\
&59552.697168&8.43029&0.00194&9.33663&46.34639&-0.00803\\
&59553.676730&13.05334&0.00250&9.34812&46.18860&-0.02586\\
&59554.576962&19.02977&0.00331&9.36551&45.58520&-0.02356\\
&59555.605078&26.50216&0.00557&9.37734&45.04094&-0.06264\\
&59557.592836&39.92466&0.00235&9.35799&45.96443&-0.03400\\
&59558.690961&45.71291&0.00203&9.33884&46.20128&-0.02282\\
&59561.597912&49.09863&0.00317&9.37203&46.28033&-0.02749\\
F&59562.588878&43.45288&0.00405&9.40179&46.07450&-0.00592\\
&59563.570834&33.56211&0.00269&9.37255&46.56322&0.00182\\
&59564.605166&20.81757&0.00255&9.35978&46.22947&-0.01380\\
F/S&59565.620260&10.95610&0.00398&9.39488&45.62137&-0.05093\\
B/F&59566.626009&7.41922&0.00318&9.39870&45.77566&-0.05742\\
&59575.684564&50.37154&0.00214&9.34620&46.22991&-0.02944\\
&59577.752093&38.32622&0.00270&9.37025&45.73892&-0.00015\\
&59580.674743&8.59139&0.00317&9.35019&45.25616&-0.02305\\
&59582.578623&11.25863&0.00264&9.33852&45.84375&-0.01116\\
&59584.601711&24.89331&0.00227&9.34412&46.15903&-0.00402\\
&59586.580398&38.52810&0.00200&9.34233&46.24767&-0.01804\\
&59588.561475&48.28672&0.00347&9.35869&45.42404&0.00701\\
&59590.718141&49.42114&0.00288&9.37818&45.52608&0.00617\\
&59590.763503&49.27649&0.00257&9.34866&45.75520&-0.03599\\
&59592.683281&34.77366&0.00219&9.34450&46.11654&-0.01601\\
&59595.638420&7.62069&0.00233&9.36676&46.04454&-0.05363\\
&59597.574782&13.47948&0.00244&9.34960&45.45612&-0.00326\\
F&59603.794993&49.94906&0.00394&9.40307&44.85825&-0.00935\\
B/F/S&59607.733856&29.58589&0.00438&9.43764&44.46652&-0.07771\\
&59610.557055&7.43040&0.00243&9.37618&45.63957&-0.00768\\
P&59613.678124&23.84446&0.00488&9.35368&45.36720&0.00387\\
&59616.558505&42.91328&0.00269&9.37147&45.71280&-0.01982\\
&59620.543282&46.73256&0.00303&9.33914&45.78754&0.02492\\
&59626.642810&12.57945&0.00368&9.33877&45.58173&-0.01760\\
&59639.611804&7.44974&0.00384&9.36696&45.70465&-0.01814\\
&59642.523956&21.14343&0.00325&9.35484&45.06757&-0.03741\\
&59645.600842&41.97985&0.00386&9.35632&46.48695&0.00010\\
S&59648.534020&50.54759&0.00650&9.36679&46.66851&-0.06742\\
&59655.659961&11.45834&0.00449&9.35032&46.25790&-0.02203\\
F&59660.602305&44.05437&0.00372&9.40772&44.79431&-0.02115\\
&59662.591082&50.49499&0.00427&9.34963&44.80618&-0.01204\\
&59664.650017&44.95555&0.00370&9.36024&45.49735&-0.02922\\
&59666.625811&23.74319&0.00842&9.38995&43.39249&-0.04636\\
&59668.519892&7.97514&0.00368&9.34184&44.44794&-0.00950\\
	\hline
\end{longtable}
\end{document}